\documentclass[twocolumn,showkeys,nofootinbib,amsmath, amssymb,aps,prd,raggedbottom]{revtex4-1}

\pdfoutput=1
 
\usepackage[utf8]{inputenc}

\usepackage{graphicx}
\usepackage{dcolumn}
\usepackage{bm}
\usepackage{color}
\usepackage{multirow}

\usepackage[colorlinks=true]{hyperref}

\newcommand{\eg}{\emph{e.g.}{} }
\newcommand{\ie}{\emph{i.e.}{} }
\newcommand{\bra}[1]{\langle #1 |}
\newcommand{\ket}[1]{| #1 \rangle}

\allowdisplaybreaks[1] 

\begin{document}

\title{New Physics in Neutron Beta Decay}
\author{Jacek~Holeczek}
\author{Micha\l{}~Ochman}
\author{El\.zbieta~Stephan}
\author{Marek~Zra\l{}ek}
\affiliation{Institute~of~Physics, University~of~Silesia, ul.~Uniwersytecka~4, 40-007~Katowice, Poland}

\date{\today}

\begin{abstract}
Limits on parameters describing physics beyond the Standard Model are presented. The most general Lorentz invariant effective Hamiltonian at the quark--lepton level involving vector, scalar and tensor operators has been used. The fits have been done using the most precise and up to date experimental data for correlation coefficients measured in free neutron beta decay as well as the Fierz term measured in superallowed Fermi decays.
\end{abstract}


\maketitle 

\section{Introduction}

The neutron decay, as the simplest beta decay, well described in the framework of the Standard Model (SM), is used as a tool to find limits on physics beyond the SM. There were many approaches to find limits on parameters describing physics beyond the SM, based on observables measured in the neutron and the nuclear beta decays, with the most recent and general ones included \eg in Refs.~\cite{Severijns,Konrad}. We decided to re-run the analysis for several reasons.  At first, most of the previous studies were based on an effective Hamiltonian at the nucleon--lepton level, which makes the distinction between New Physics and nucleon structure not so clear. Secondly, we aim to find bounds on the parameters which in future will be particularly useful in the description of neutrino states in beta beam oscillation experiments.
Moreover, our new fits use the most precise and up to date experimental data from correlation coefficients $a$, $A$, $B$ measured in free neutron beta decay (see TABLE~\ref{data_table}) and the Fierz term measured in superallowed Fermi decays (the $b_F$, see Ref.~\cite{Hardy_Towner}).
This work continues and extends our previous researches from Ref.~\cite{our_acta_2011}.

Similarly as in Ref.~\cite{Herczeg} we would like to present an analysis starting at the quark--lepton level, with a particularly simple distinction between the New Physics parameters and the nucleon form factors in the vector part of the interaction. However, in contrary to what was done in Ref.~\cite{Herczeg}, we performed a self-consistent least squares analysis to the free neutron decay data, putting careful attention to the still unknown value of the Fierz term $b$.
Our paper is divided, except for this short introduction, into five sections. In Section~\ref{sec:formalism} we give a short introduction into the beta decay formalism needed for our purposes. In Section~\ref{sec:parameters} we show the definitions of the fitted parameters and in Section~\ref{sec:analysis} the details of the data analysis are given. In Section~\ref{sec:results} we present our results and the conclusions are summarized in Section~\ref{sec:conclusions}. In the Appendix, the formulas for decay parameters in the free neutron decay are presented.

\section{\label{sec:formalism}Formalism of the Neutron Beta Decay}

The most general, derivative--free and Lorentz invariant four fermion contact interaction for neutron decay at nucleon--lepton level was first introduced by Lee and Young \cite{Lee} and later used in the canonical work by Jackson, Treiman and Wyld \cite{Jackson} to obtain formulas for complete set of decay parameters. Such general interaction involving scalar, vector and tensor terms can be parametrized in many ways (not only as in Refs.~\cite{Lee,Jackson}). Herein, we will use the parametrization introduced in Ref.~\cite{Herczeg} with the interaction Hamiltonian at the quark--lepton level in form
\begin{eqnarray} \label{hamiltonian}
\mathcal{H}_{\beta} & = & 4 \sum_{i=1}^{n} \sum_{k,l = L,R}
\biggl\{ 
 A_{kl}  U^{k}_{ei}  \bar{e} P_k \nu_i  \bar{u} P_l d  \nonumber \\
& & +
 a_{kl}  U^{k}_{ei}  \bar{e} \gamma_\mu P_k \nu_i  \bar{u} \gamma^\mu P_l d  \nonumber \\
& & +
 \alpha_{kk}  U^{k}_{ei}  \bar{e} \frac{\sigma_{\mu\nu}}{\sqrt{2}} P_k \nu_i  
\bar{u} \frac{\sigma^{\mu\nu}}{\sqrt{2}} P_k d 
\biggl\}+\textnormal{H.c.,}
\end{eqnarray}
where $P_L = \frac{1}{2}\left( 1 - \gamma_5 \right)$, $P_R = \frac{1}{2}\left( 1 + \gamma_5 \right)$, $\sigma_{\mu\nu} = \frac{i}{2}\left[\gamma_\mu,\gamma_\nu\right]$ (the metric and gamma matrices are the same as \eg in Ref.~\cite{Giunti}), $u$ and $d$ are quark fields, $e$ stands for the electron field and $\nu_i$ is the $i$-th neutrino field with a certain mass, $U^L$ and $U^R$ are $3 \times n$ mixing matrices\footnote{Comparing the notation used in Ref.~\cite{Herczeg} we substitute $U$ with $U^L$ and $V$ with $U^R$.} for the Left and Right--handed neutrinos respectively (we work in the basis in which mass matrix of charged leptons is diagonal). We assume in our work that $a_{kl}$, $A_{kl}$, $\alpha_{kk}$ for $k,l = L,R$ are real, contrary to Ref.~\cite{Herczeg}. The SM is restored when $n=3$ (then $U^L$ is the well known Pontecorvo--Maki--Nakagawa--Sakata matrix), $a_{kl}=0$, $A_{kl}=0$, $\alpha_{kk}=0$ for $k,l = L,R$ except $a_{LL} = a_{LL}^{SM} = V_{ud} G_F/\sqrt{2}$, where $G_F$ is the usual Fermi constant and $V_{ud}$ is the element of quark Cabibbo--Kobayashi--Maskawa mixing matrix.

When calculating amplitudes for the neutron beta decay we used the following relations \cite{Herczeg}
\begin{subequations}
\begin{eqnarray}
g_{V}\bar{u}_{p} \gamma_\mu u_{n} & = & \bra{p} \bar{u}
\gamma_\mu d \ket{n}\textnormal{,}  \label{g_V_def} \\
g_{A}\bar{u}_{p} \gamma_\mu \gamma_5 u_{n} & = & \bra{p}
\bar{u} \gamma_\mu \gamma_5 d \ket{n}\textnormal{,}  \\
g_{S}\bar{u}_{p} u_{n} & = & \bra{p} \bar{u} d \ket{n}\textnormal{,}  \\
g_{T}\bar{u}_{p} \sigma_{\mu\nu} u_{n} & = & 
\bra{p} \bar{u} \sigma_{\mu\nu} d \ket{n}\textnormal{,} \label{g_T_def}
\end{eqnarray}
\end{subequations}
where $g_i = g_i(q^2 \approx 0)$ for $i = V,A,S,T$ are the form factors at small four-momentum transfer $q = p_p - p_n$ and $\bra{p}$, $p_p$, $u_{p}$ ($\ket{n}$, $p_n$, $u_{n}$) are the proton (neutron) state, four-momentum and the ordinary free-particle Dirac wave function. From conserved vector current hypothesis, neglecting small corrections, we know that $g_{V} = 1$ (see Ref.~\cite{Severijns} for a review). There were many attempts to calculate $g_A$ and $g_T$ in lattice QCD, which were summarized and supplemented with the first calculation of $g_S$ in Ref.~\cite{neutron_BSM_long} as follows: (i)~the central value of $g_A$ is in the range $1.12 < g_A < 1.26$ and when we include the errors (that are not competitive with the experimental uncertainty on the $g_A/g_V$ ratio --- see section~\ref{sec:results}) we can conclude that $g_A$ can be roughly in the range from $1.1$ to $1.34$, (ii)~the values of form factors beyond SM are $g_S = 0.8 \pm 0.4$ and $g_T = 1.05 \pm 0.35$. In the later derivations in this work $g_{S}$, $g_{T}$, $g_V$ and $g_A$ are treated as free real parameters (we assume time reversal invariance of strong interactions).

\section{\label{sec:parameters}Decay Parameters}

Using Eq.~(\ref{hamiltonian}) we calculated the five-fold differential decay width for polarized neutron in its rest frame without measurement of final electron and proton polarizations. As in Ref.~\cite{Jackson}, in our calculations we have neglected QED corrections (unless otherwise stated) and recoil effects (for a review see \eg Ref.~\cite{neutron_BSM_long}), obtaining
\begin{eqnarray} \label{d_Gamma}
\frac{d\Gamma}{dE_e d\Omega_e d\Omega _\nu}
& = & \frac{p_e E_e E_\nu^2}{(2\pi)^5} 
G_\beta \xi \bigg\{   
1 + a \frac{\Vec{p}_e \cdot \Vec{p}_\nu}{E_e E_\nu} + 
b \frac{m_e}{E_e} \nonumber \\
& & 
 + \Vec{\lambda}_n \cdot \left[ A
\frac{\Vec{p}_e}{E_e} + B \frac{\Vec{p}_\nu}{E_\nu} + D
\frac{\Vec{p}_e \times \Vec{p}_\nu}{E_e E_\nu} \right]
\bigg\} \textnormal{,} \nonumber \\
& & 
\end{eqnarray}
where $\Omega_e$ and $\Omega _\nu$ denote the solid angles of electron and anti-neutrino emission, $m_e$, $\Vec{p}_e$, $E_e$ are, respectively, the mass, momentum ($p_e = |\Vec{p}_e|$) and total energy of an electron, $\Vec{p}_\nu$ and $|\Vec{p}_\nu| = E_\nu = E_0 - E_e$ are the anti-neutrino momentum and energy\footnote{The effect of nonzero neutrino masses enters only through presence of mixing matrices $U^L$ and $U^R$.}, $E_0 = m_n - m_p$ (that is a difference between neutron and proton masses) is the maximum value of $E_e$ and $\Vec{\lambda}_n$ is the neutron polarization vector. 

The value of $G_\beta = 2 (a_{LL}+a_{LR})^2 g_V^2 \sum_i' | U^L_{ei} |^2$ (the prime in the sum indicates that the summation runs only over kinematically allowed anti-neutrino states) can be accessed through measurements of the total decay width --- it is however beyond our present interest. The $D$ correlation coefficient is mentioned here only for completeness, since $D \equiv 0$ upon approximations under considerations and for real parameters (experimentally one has \cite{PDG}: $D=(-1.2 \pm 2.0) \times 10^{-4}$). The $B$ correlation coefficient has the form of
\begin{equation} \label{B_coeff}
B = B_{0} + b_\nu\frac{ m_e}{E_e},
\end{equation}
where the formulas for $B_{0}$ and $b_\nu$ as well as for the correlation coefficients $a$, $b$, $A$ and the common factor $\xi$ are listed in the Appendix --- they are expressed in terms of the following parameters (similar as in Ref.~\cite{Herczeg}) for $k,l = L,R$
\begin{subequations} \label{VST_parameters}
\begin{eqnarray}
\lambda & = &\frac{g_A}{g_V}\frac{\bar{a}_{LL}-\bar{a}_{LR}}{\bar{a}_{LL}+\bar{a}_{LR}} , \label{lambda_def}\\
 &  & \nonumber \\
V_{RL}  & = & \frac{\bar{a}_{LL} \bar{a}_{RL}-\bar{a}_{LR} \bar{a}_{RR}}{\bar{a}_{LL}^2-\bar{a}_{LR}^2} ,\\
 &  & \nonumber \\
V_{RR}  & = & \frac{\bar{a}_{LL} \bar{a}_{RR}-\bar{a}_{LR} \bar{a}_{RL}}{\bar{a}_{LL}^2-\bar{a}_{LR}^2} ,\\
 &  & \nonumber \\
S_{kl}   & = & \frac{g_S}{g_V} \frac{\bar{A}_{kl}}{\bar{a}_{LL}+\bar{a}_{LR}} ,\\
 &  & \nonumber \\
T_{kk}  & = & \frac{g_T}{g_V} \frac{\bar{\alpha}_{kk}}{\bar{a}_{LL}+\bar{a}_{LR}} ,
\end{eqnarray}
\end{subequations}
where\footnote{Note that $V_{RR} + V_{RL} = (\bar{a}_{RR}+\bar{a}_{RL})/(\bar{a}_{LL}+\bar{a}_{LR})$ and $V_{RR} - V_{RL} = (\bar{a}_{RR}-\bar{a}_{RL})/(\bar{a}_{LL}-\bar{a}_{LR})$.}
\begin{subequations} \label{akl_def}
\begin{eqnarray} 
\bar{a}_{kl} & = & a_{kl} \sqrt{\textstyle \sum_i' | U^{k}_{ei} |^2}, \\
\bar{A}_{kl} & = & A_{kl} \sqrt{\textstyle \sum_i' | U^{k}_{ei} |^2}, \\
\bar{\alpha}_{kk} & = & \alpha_{kk}  \sqrt{\textstyle \sum_i' | U^{k}_{ei} |^2}
\end{eqnarray}
\end{subequations}
with summation running only over kinematically allowed anti-neutrino states. For $a_{LR} \equiv 0$ it is easy to separate the dependency of decay parameters on the ratio of nucleon form factors $\lambda = g_A/g_V$ from the rest of parameters.

For scalar couplings, in general, the formulas for correlation coefficients (see the Appendix) depend only on two combinations
\begin{subequations} \label{s_L_R}
\begin{eqnarray}
s_L & = & S_{LL} + S_{LR}\textnormal{,}\\
s_R & = & S_{RR} + S_{RL}\textnormal{.} 
\end{eqnarray}
\end{subequations}

Note that setting $\bar{a}_{LR} = 0$ results in $V_{Rk} = \bar{a}_{Rk} / \bar{a}_{LL}$ for $k = L,R$ and our previous parametrization \cite{our_acta_2011} is restored, except for the old $V_{LR} = \bar{a}_{LR} / \bar{a}_{LL}$ parameter which no longer appears as the $\bar{a}_{LR}$ explicitly enters our present parametrization --- see  Eqs.~(\ref{VST_parameters}) (note also the small change of notation $U = U^L$ and $V = U^R$). In fact, even for $\bar{a}_{LR} \neq 0$ the results presented in plots in Ref.~\cite{our_acta_2011} can be interpreted in terms of our present parametrization provided that $g_A/g_V$ in those plots is replaced with $\lambda$, as defined in Eq.~(\ref{lambda_def}).

\section{\label{sec:analysis}Least Squares Analysis}

Following the approach applied in Refs.~\cite{Gluck,Severijns,Konrad}, the measurements of $a$, $A$, $B$ are interpreted as measurements of the respective quantities
\begin{subequations}
\begin{eqnarray}
\bar{a}( \langle W^{-1}\rangle ) & = & \frac{ a }{ 1 + b \langle W^{-1}\rangle } \textnormal{,} \\	
\bar{A}( \langle W^{-1}\rangle ) & = & \frac{ A }{ 1 + b \langle W^{-1}\rangle } \textnormal{,} \\
\bar{B}( \langle W^{-1}\rangle ) & = & \frac{ B_{0} + b_\nu \langle W^{-1}\rangle }
                                                    { 1 + b \langle W^{-1}\rangle } 
\textnormal{,} 
\end{eqnarray}
\end{subequations}
where $\langle W^{-1}\rangle = m_e\langle E_e^{-1} \rangle$. This procedure arises mainly from the fact that experimentalists analyze their data assuming $b \equiv 0$ and $b_\nu \equiv 0$. Note that in the SM and for some combinations of parameters of physics beyond SM this assumption is valid.

The $\chi^2$, which is minimized with the fit procedure, is of the form
\begin{eqnarray} \label{chi2}
\chi^2 	& = &   \sum_i \left[\frac{a_i - \bar{a}(\langle W^{-1}\rangle_i)}{\delta a_i}\right]^2
				\nonumber \\
		& + &   \sum_j \left[\frac{A_j - \bar{A}(\langle W^{-1}\rangle_j)}{\delta A_j}\right]^2 
				\nonumber \\
		& + &   \sum_k \left[\frac{B_k - \bar{B}(\langle W^{-1}\rangle_k)}{\delta B_k}\right]^2 
				\textnormal{,}
\end{eqnarray}
where the selected data are presented in TABLE~\ref{data_table}: $a_i$, $A_j$, $B_k$ and $\delta a_i$, $\delta A_j$, $\delta B_k$ denote the central value and the error of the respective decay parameter in a certain experiment. We follow the PDG \cite{PDG} data selection, but (i)~we have used the corrected value for Ref.~\cite{LIU_10} given in Ref.~\cite{MENDENHALL_12}, (ii)~we added new measurements of $A$ \cite{MENDENHALL_12,MUND_12} and dropped older measurements of this decay parameter \cite{LIAUD_97,YEROZOLIMSKY_97,BOPP_86} since they are poorly consistent with the newer ones and (iii)~we kept only the most precise measurements of $a$ and $B$ ($\delta a_i /a_i \leq 6\%$ and $\delta B_k/B_k \leq 2\%$). When both statistical and systematic errors were reported separately, these two errors were added in quadrature. In the case of asymmetric errors the larger of the reported errors was taken.

\begin{table}
\centering
\begin{tabular}{cllllll}
PAR.  &  VALUE  &  ERROR  &  $\langle W^{-1} \rangle$  &  PAPER ID &  &   \\ \hline
  &     &     &     &     &  &   \\  
$a$  &  $-0.1054$  &  $0.0055$  &  $0.655$  &  BYRNE      & 02  &  \cite{BYRNE_02}  \\
 &  $-0.1017$  &  $0.0051$  &  $0.655$  &  STRATOWA   & 78   &  \cite{STRATOWA_78}  \\
 &       &      &      &             &      &  \\
$A$ &  $-0.11954$  &   $0.00112$  &   $0.576$  &  MENDENHALL & 12   &  \cite{MENDENHALL_12}  \\
 &  $-0.11996$  &   $0.00058$  &   $0.553$  &  MUND       & 12   &  \cite{MUND_12}  \\
 &  $-0.11942$  &   $0.00166$  &   $0.557$  &  LIU        & 10   &  \cite{LIU_10,MENDENHALL_12}  \\
 &  $-0.1189$  &  $0.0007$  &  $0.534$  &  ABELE      & 02   &  \cite{ABELE_02}  \\
 &       &      &      &             &      &  \\
$B$  &  $0.980$  &  $0.005$  &  $0.599$  &  SCHUMANN   & 07   &  \cite{SCHUMANN_07}  \\
 &  $0.967$  &  $0.012$  &  $0.600$  &  KREUZ      & 05   &  \cite{KREUZ_05}  \\
 &  $0.9801$  &  $0.0046$  &  $0.594$  &  SEREBROV   & 98   &  \cite{SEREBROV_98}  \\
 &  $0.9894$  &  $0.0083$  &  $0.554$  &  KUZNETSOV  & 95   &  \cite{KUZNETSOV_95}  \\
\end{tabular}
\caption{\label{data_table}The experimental values of correlation coefficients measured in free neutron beta decay. All PAPER ID names were taken from PDG \cite{PDG} except the new ones for Ref.~\cite{MENDENHALL_12} and Ref.~\cite{MUND_12}.}
\end{table}

Part of the values of $\langle W^{-1}\rangle_i = m_e\langle E_e^{-1} \rangle_i$, present in TABLE~\ref{data_table}, were taken from Ref.~\cite{Severijns}. The remaining ones were calculated using
\begin{equation} \label{Gamma_Fermi}
\langle E_e^{-1} \rangle_i = 
{\displaystyle \int_{E_i^{min}}^{E_i^{max}} d E_e \frac{d\Gamma_{SM}}{dE_e} E_e^{-1}} \bigg/
       {\displaystyle \int_{E_i^{min}}^{E_i^{max}} d E_e \frac{d\Gamma_{SM}}{dE_e}}
  \textnormal{,}
\end{equation}
where $E_i^{min}$ and $E_i^{max}$ are the limits of the energy accepted in the experiment. The differential decay width was calculated within SM, therefore (compare \eg Ref.~\cite{on_continuum})
\begin{equation}
  \frac{d\Gamma_{SM}}{dE_e}=(g_V^2 + 3 g^2_A) \frac{G^2_F |V_{ud}|^2}{2
    \pi^3} p_e E_e (E_0 - E_e)^2 F(E_e)    \textnormal{,}
\end{equation}
where the Fermi function $F(E_e)$, that is the leading order QED correction, was approximated (for an exact calculation see Ref.~\cite{Fermi}) as in \eg Ref.~\cite{Schopper} to the form of
\begin{equation}
F(E_e)=\frac{2\pi\alpha E_e / p_e}
{\displaystyle 1 - e^{\textstyle -2\pi \alpha
E_e/p_e}}   \textnormal{.}
\end{equation}

The $V_{RL}$, $V_{RR}$, $s_R$, $T_{RR}$ enter quadratically or as mixed terms between pairs of these couplings in the formulas for the correlation coefficients (see the Appendix), whereas $s_L$ and $T_{LL}$ enter also linearly. Therefore, the $\chi^2$ function in Eq.~(\ref{chi2}) reveals the symmetry
\begin{eqnarray} \label{chi2_symmetry}
& \chi^2(\lambda, s_L, T_{LL}, V_{RL}, V_{RR}, s_R, T_{RR}) = \nonumber \\
& \chi^2(\lambda, s_L, T_{LL}, -V_{RL}, -V_{RR}, -s_R, -T_{RR}) \textnormal{.}
\end{eqnarray}
The linear terms appear only in $b$ and $b_\nu$ so these quantities are of special importance from the point of view of setting limits on $s_L$ and $T_{LL}$.

\section{\label{sec:results}Results}

\paragraph{The $\lambda$ value.}

Let us first consider the case when all $a_{kl}$, $A_{kl}$, $\alpha_{kk}$ parameters are zero except $a_{LL}$ and $a_{LR}$. This results in $V_{RL} = 0$, $V_{RR} = 0$ as well as $T_{kk} = 0$ and $s_k = 0$ for $k = L,R$ so that the only nonzero parameter is $\lambda$ given in Eq.~(\ref{lambda_def}). In this case $b \equiv 0$ as well as $b_\nu \equiv 0$ and the formulas for decay parameters simplify to the well known SM expressions
\begin{subequations}
\begin{eqnarray}
a & = & -\frac{\lambda^2-1}{3\lambda^2+1} \textnormal{,} \\
A & = & -\frac{2\lambda(\lambda-1)}{3\lambda^2+1} \textnormal{,} \\
B & = & \frac{2\lambda(\lambda+1)}{3\lambda^2+1} \textnormal{.}
\end{eqnarray}
\end{subequations}
whereas $\lambda$ can go beyond its SM form of $\lambda = g_A/g_V$ (see Eq.~(\ref{lambda_def}), note that in our convention $\lambda > 0$ contrary to \eg PDG \cite{PDG}).
In this case, the one-parameter fit to the data presented in the TABLE~\ref{data_table} is performed, which results in $\chi^2_{min} = 9.542$ (the value of $\chi^2$ at minimum) with
\begin{equation} \label{g_A_fit}
\lambda = 1.2755 \pm \left\{
\begin{array}{l} 
0.0011\ (68.27\%\ \textnormal{C.L.}),	\\
0.0018\ (90\%\ \textnormal{C.L.}),	\\
0.0022\ (95.45\%\ \textnormal{C.L.}).
\end{array}
\right.
\end{equation}
The PDG average is \cite{PDG}: $\lambda = 1.2701 \pm 0.0025$ (the error was scaled by PDG by $1.9$, we changed the sign to meet our convention) which differs from our value because of different data selection, mainly of the $A$ decay parameter. 

\paragraph{Many--parameter fits.}

\begin{figure*}[p!]
\centering

\includegraphics[width=0.32\textwidth]{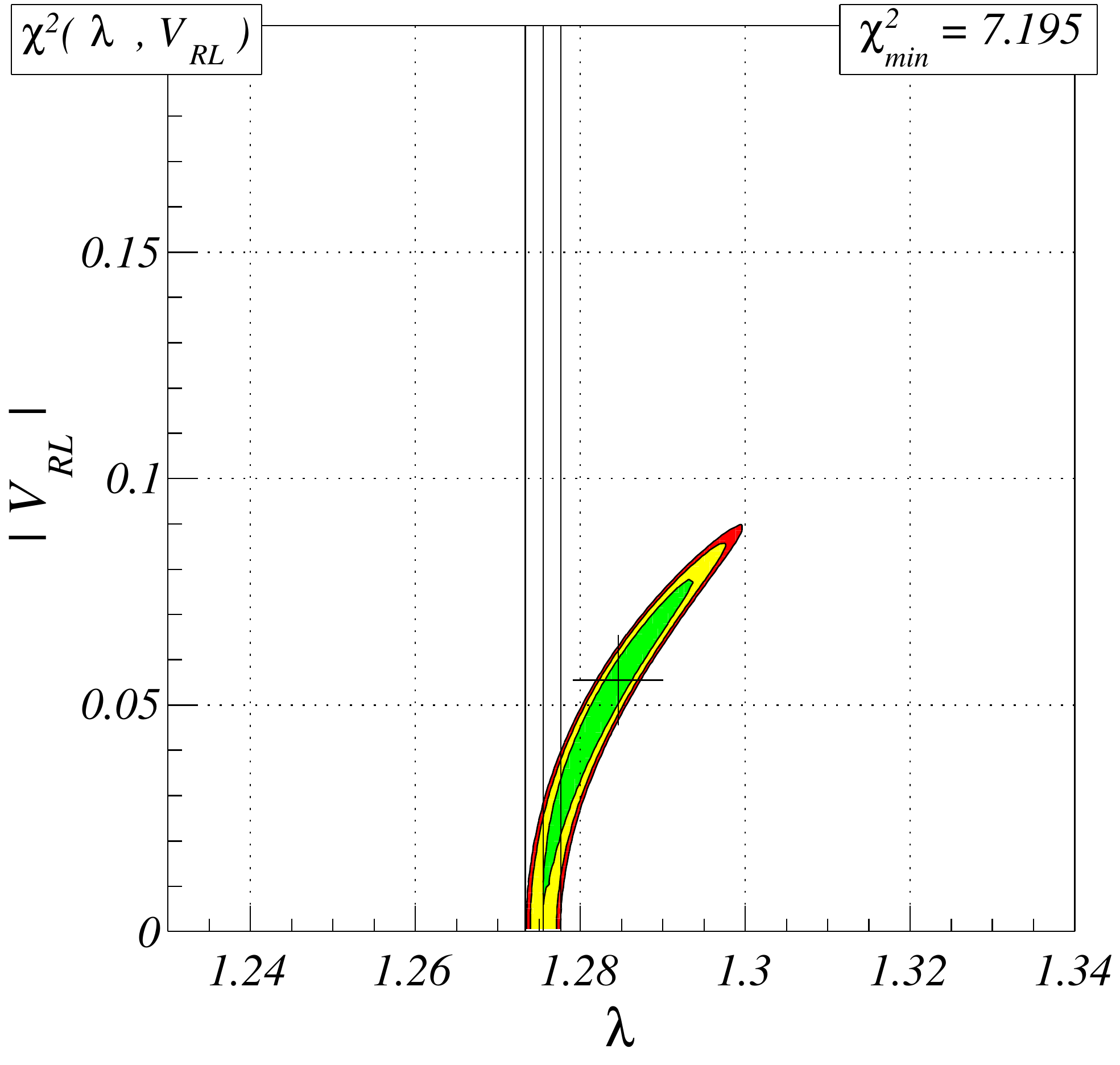}
\includegraphics[width=0.32\textwidth]{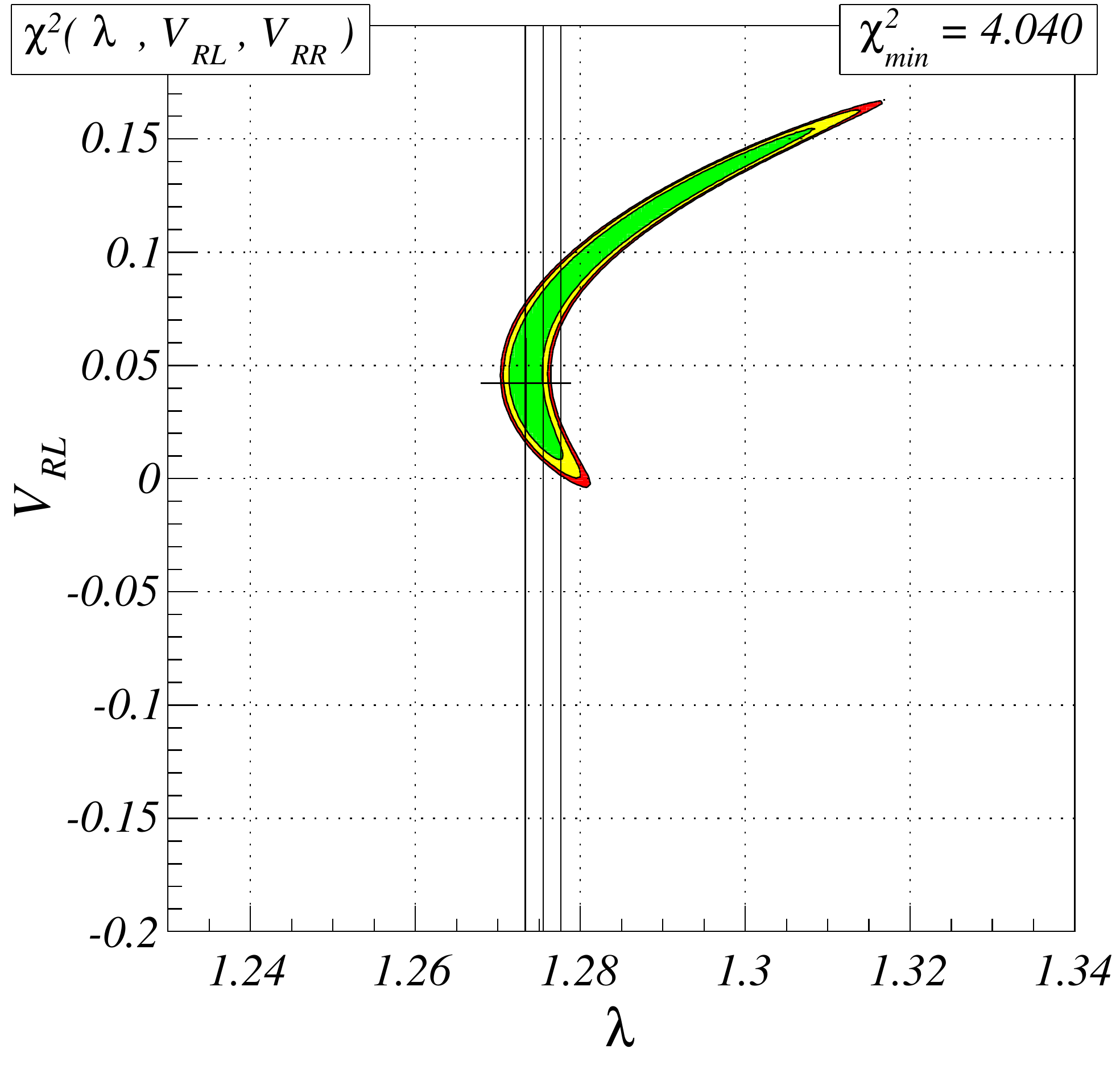}
\includegraphics[width=0.32\textwidth]{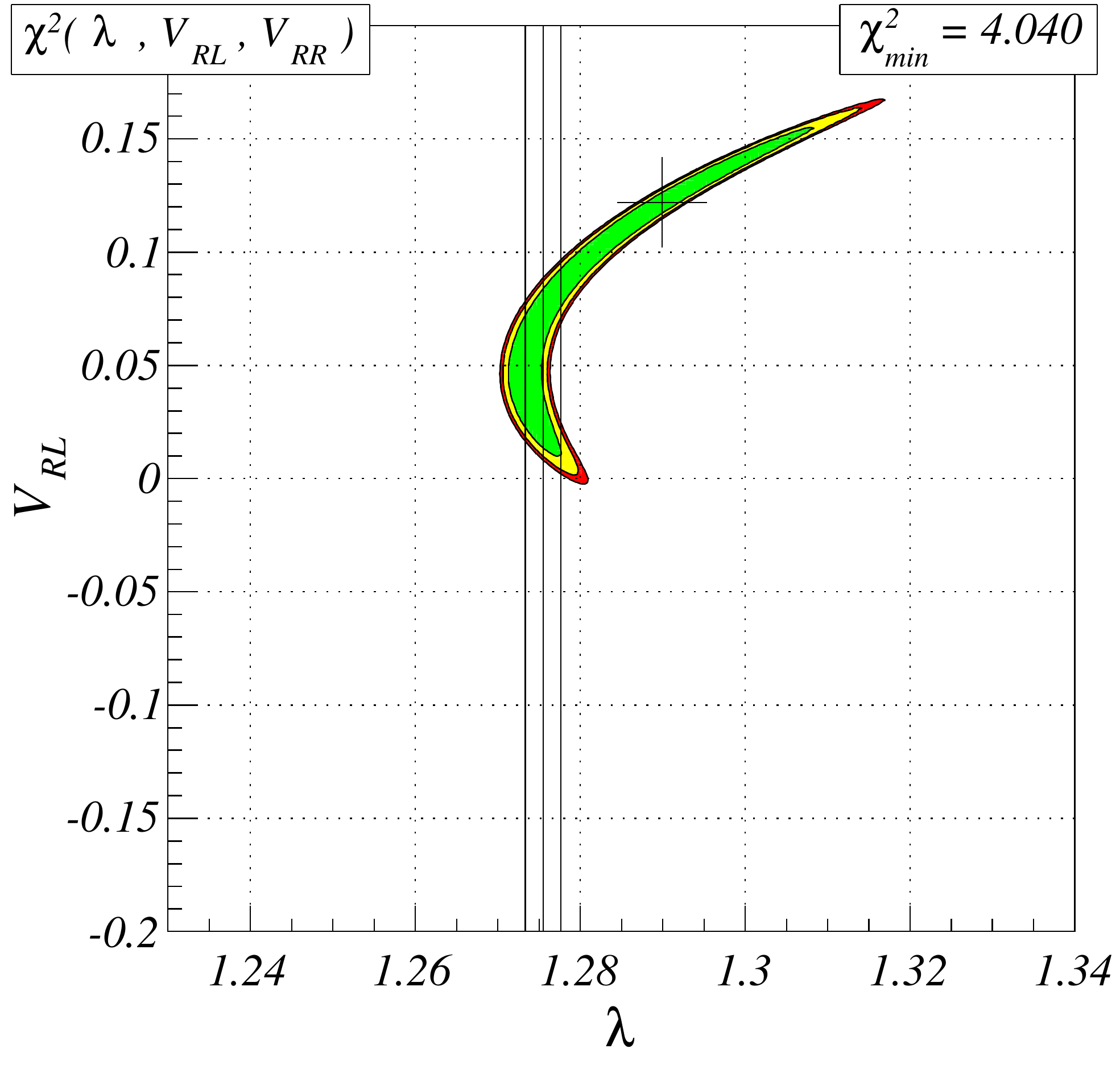}

\includegraphics[width=0.32\textwidth]{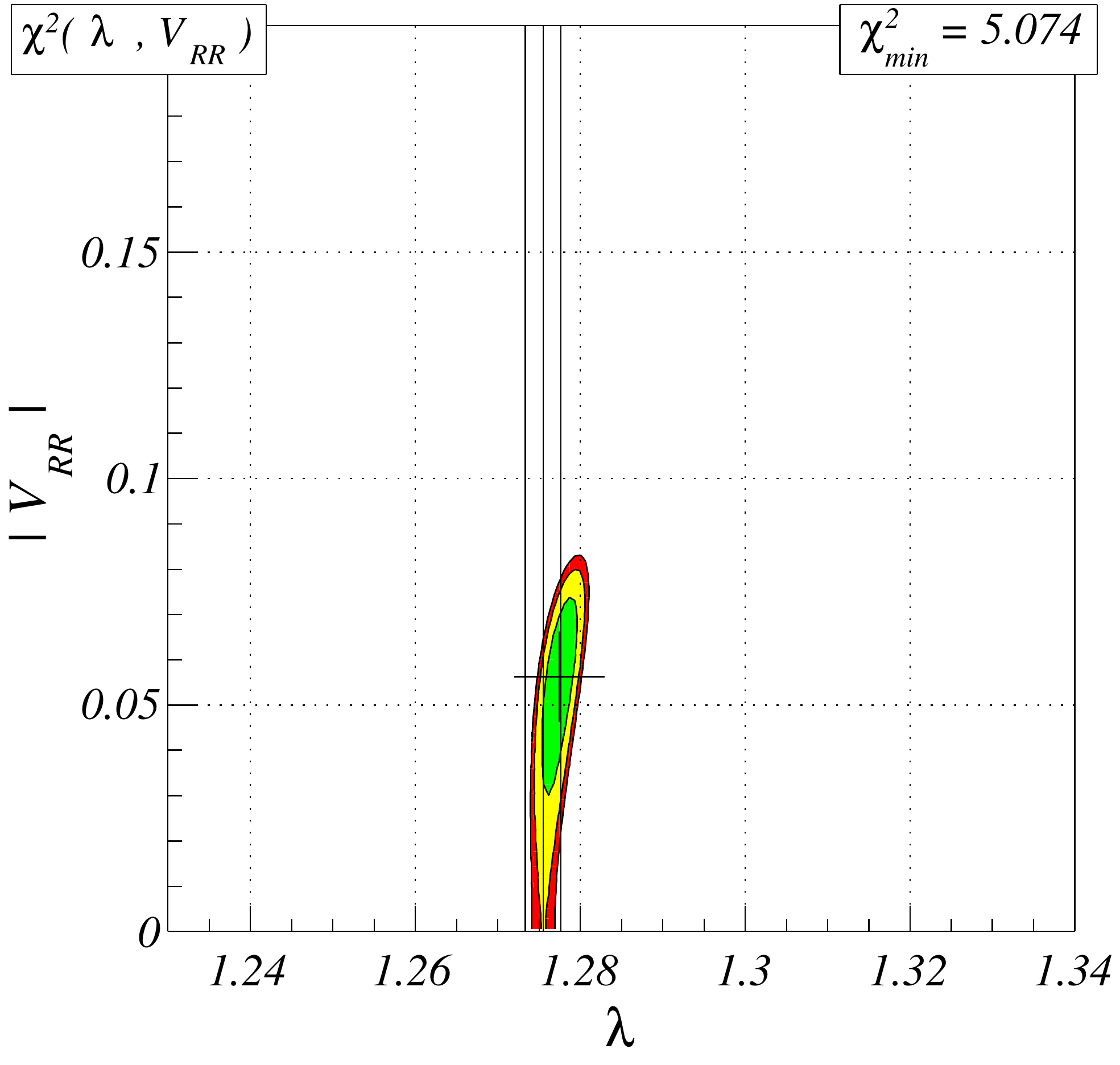}
\includegraphics[width=0.32\textwidth]{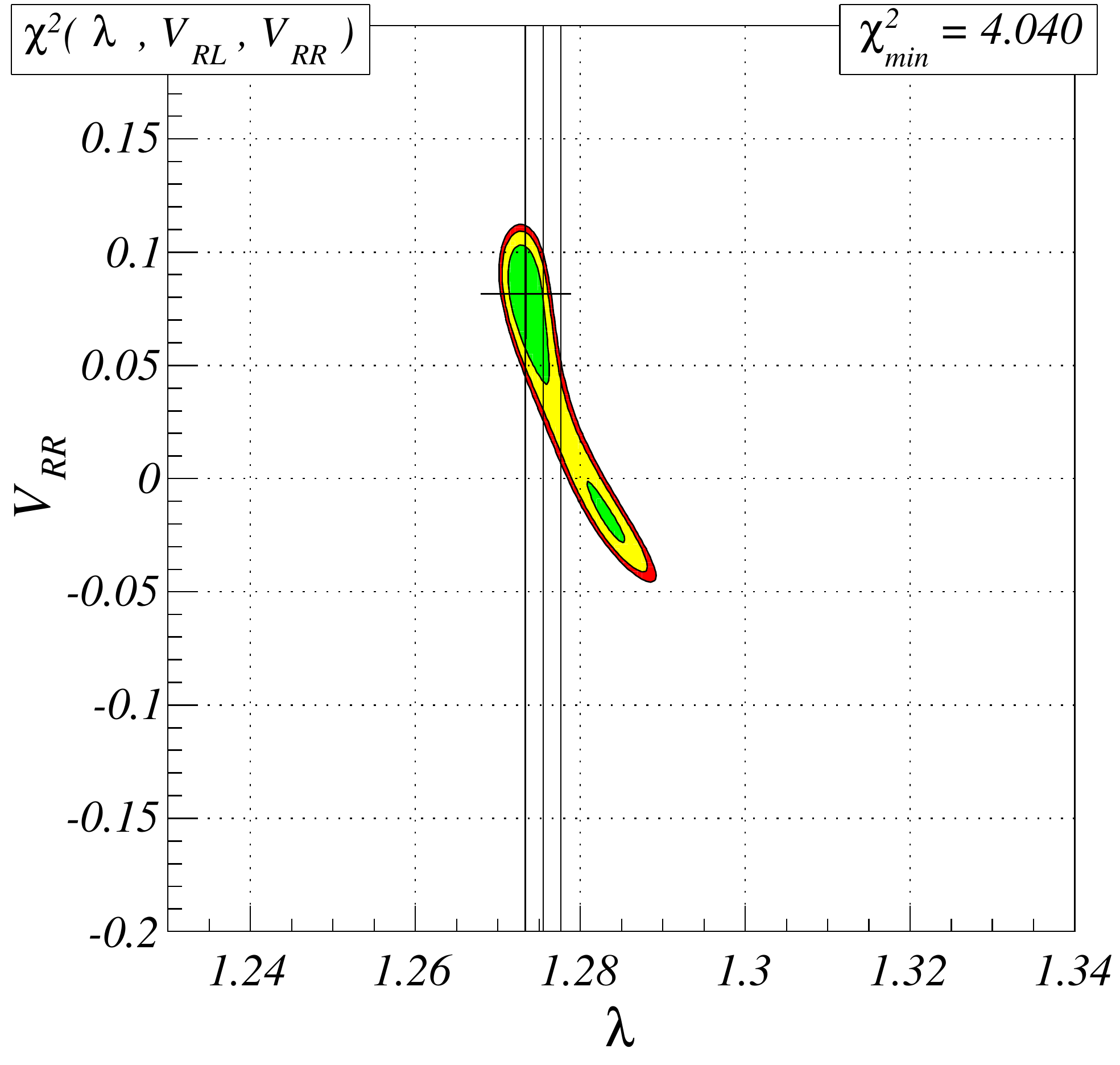}
\includegraphics[width=0.32\textwidth]{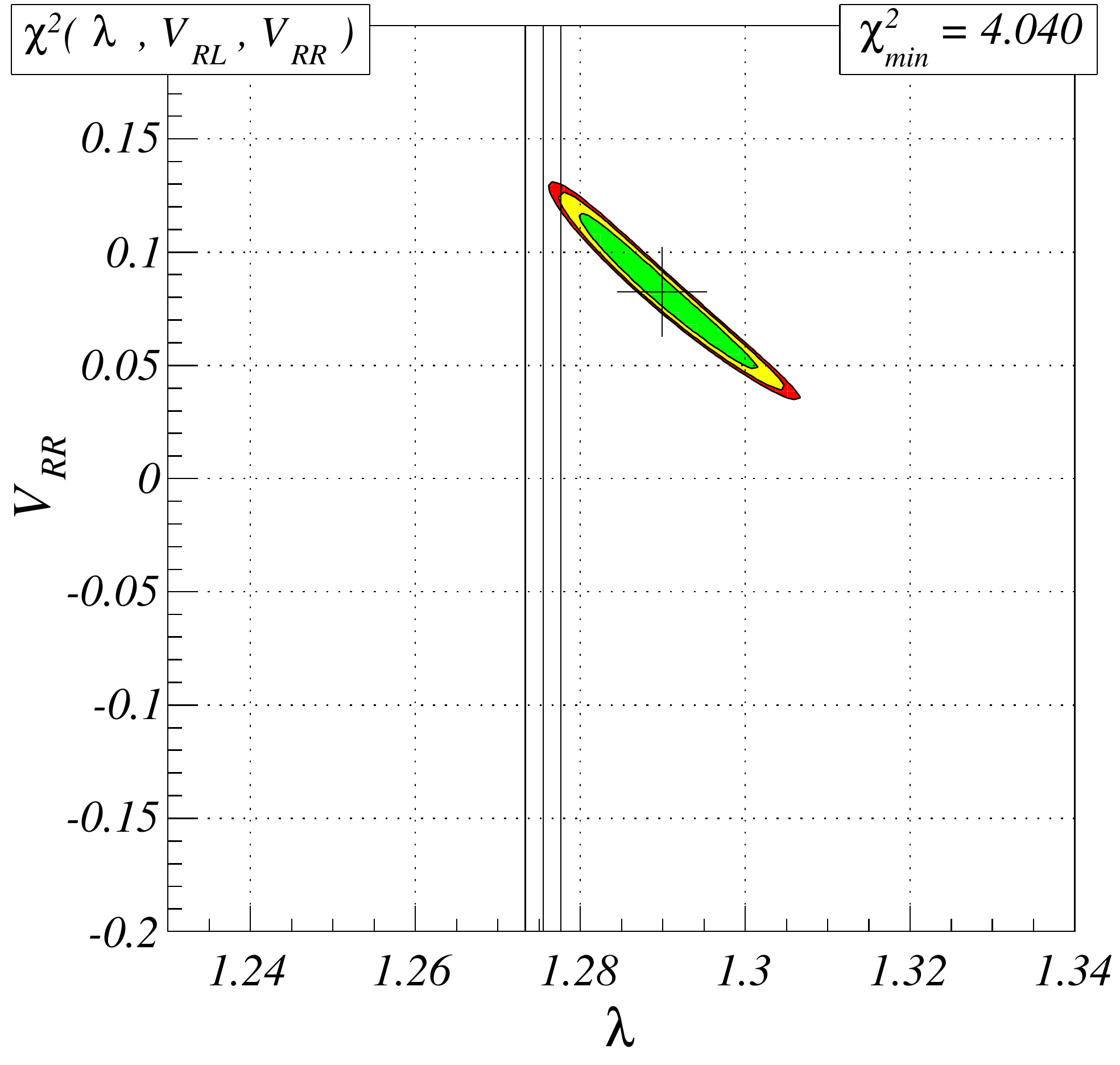}

\includegraphics[width=0.32\textwidth]{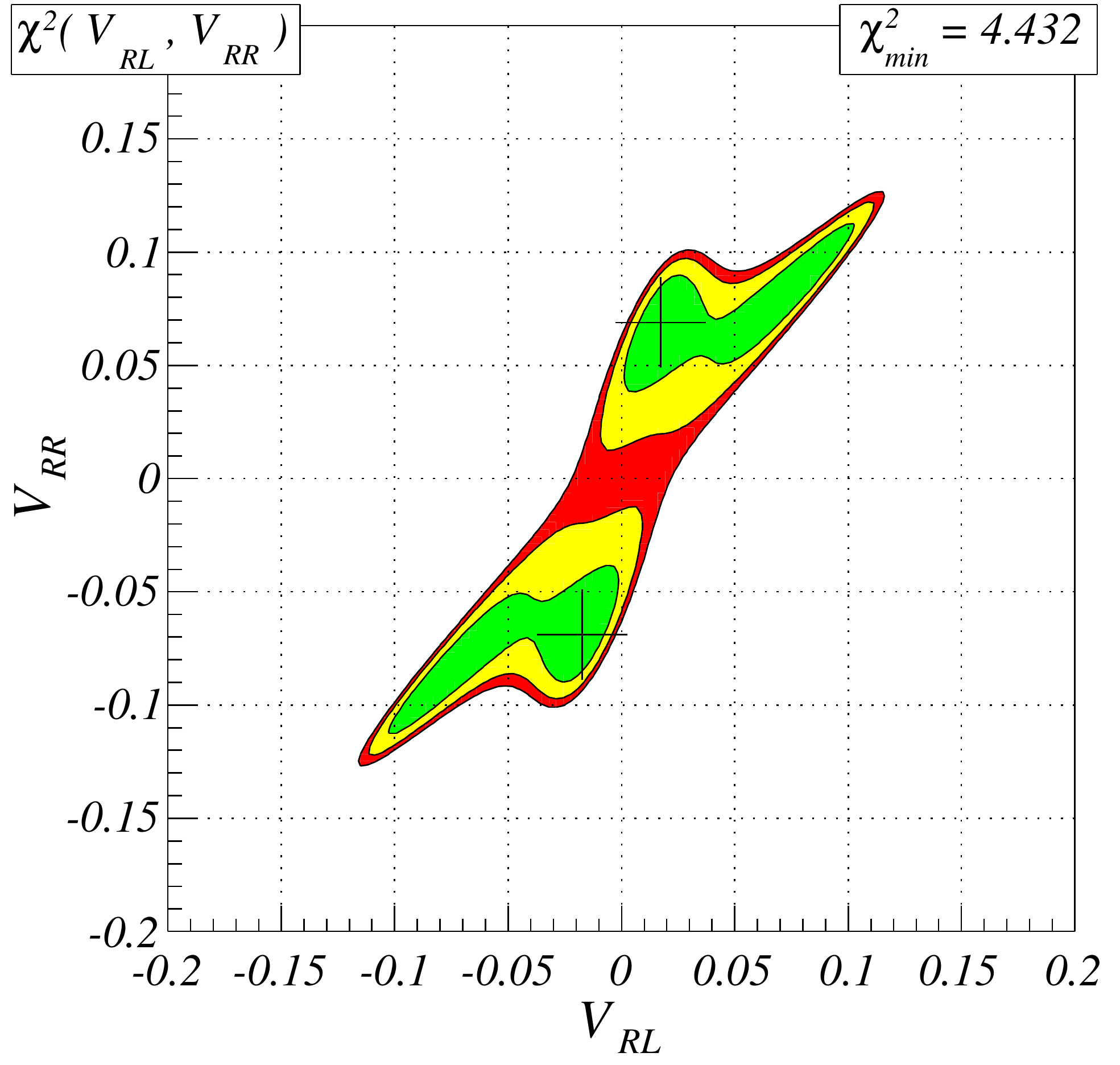}
\includegraphics[width=0.32\textwidth]{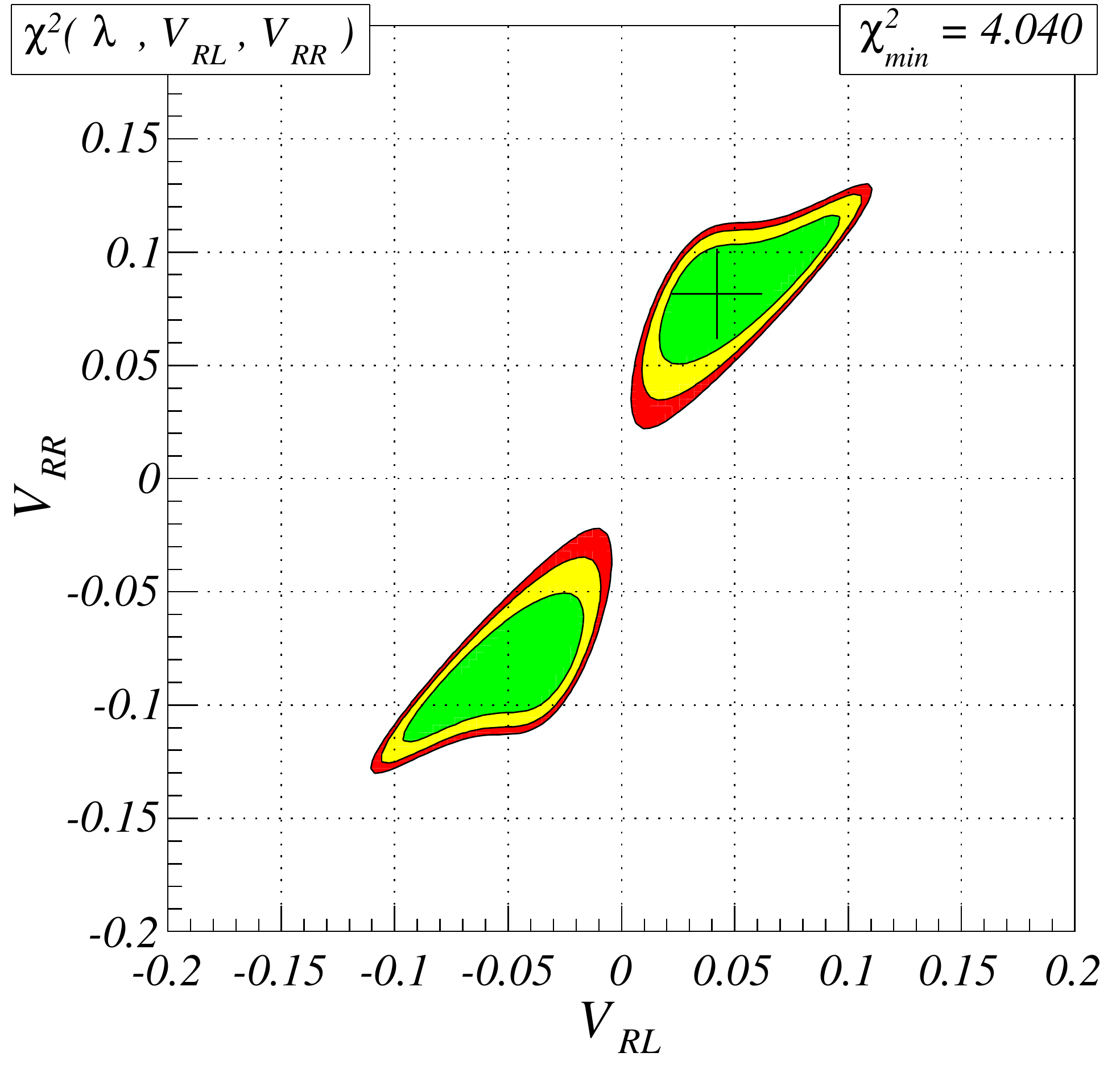}
\includegraphics[width=0.32\textwidth]{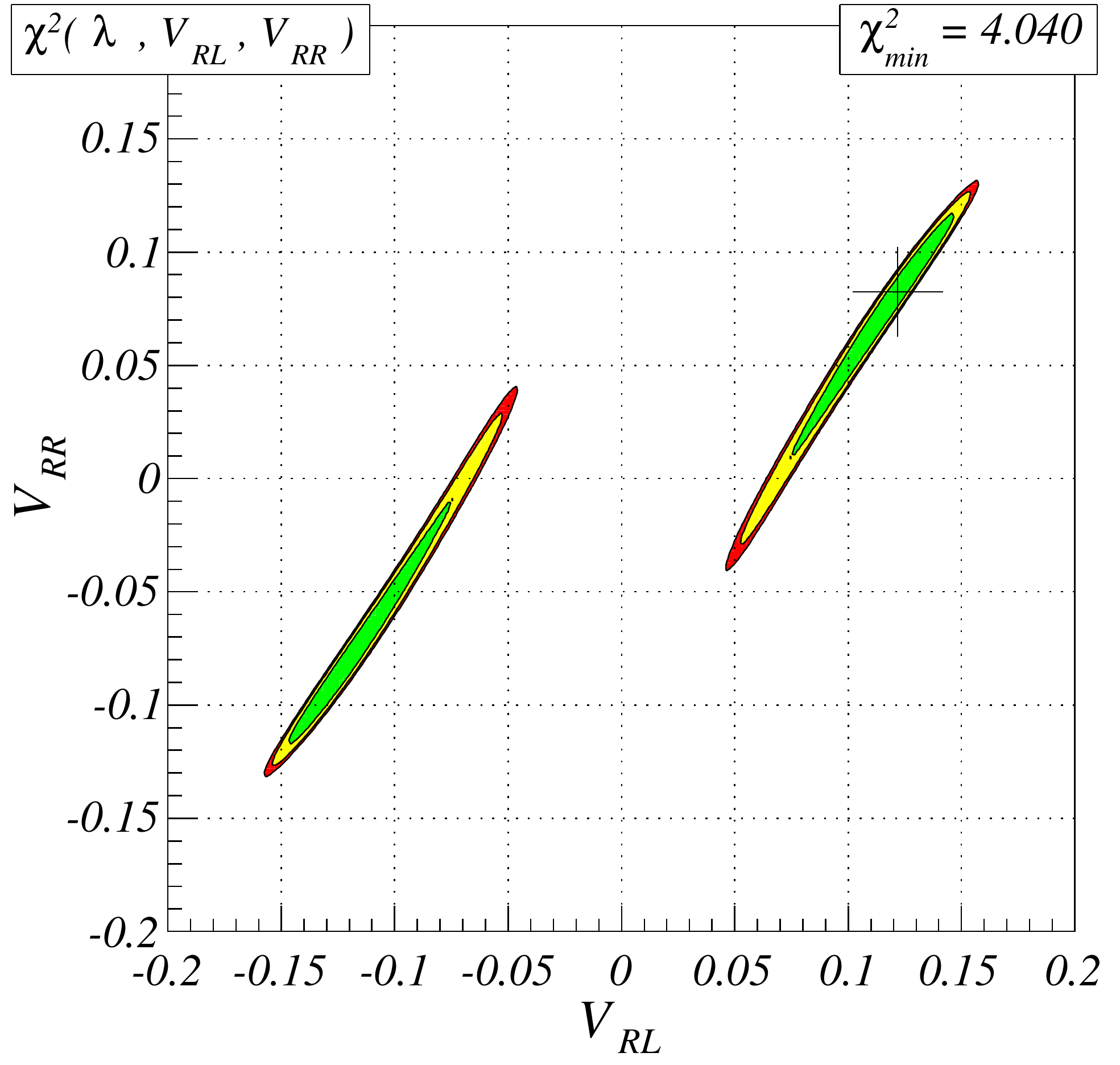}

\includegraphics[width=0.5\textwidth]{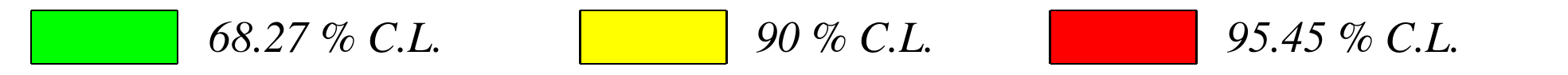}

\caption{\label{V_RL_V_RR}The results of fits of vector couplings to the data presented in the TABLE~\ref{data_table}. The list of fitted parameters is indicated as the arguments of the $\chi^2$ function in the left upper corner of each plot and the value of the $\chi^2$ at minimum is shown in the right upper corner. The cross marks the position of the $\chi^2$ minimum. The solid vertical lines mark areas that correspond to the $95.45\%\ \textnormal{C.L.}$ interval of the one-parameter fit on $\lambda$ (Eq.~(\ref{g_A_fit})). In the left column, results of two-parameter fits are shown. In the center and right columns we present slices of the three-dimensional $\chi^2$ volume in two of four equivalent minima of the respective three-parameter fit --- the minima corresponding to different values of $\lambda$ are in separate columns. The slices in the remaining two minima can easily be obtained through appropriate symmetries according to Eq.~(\ref{chi2_symmetry}).}
\end{figure*}

\begin{figure*}[p!]
\centering

\includegraphics[width=0.32\textwidth]{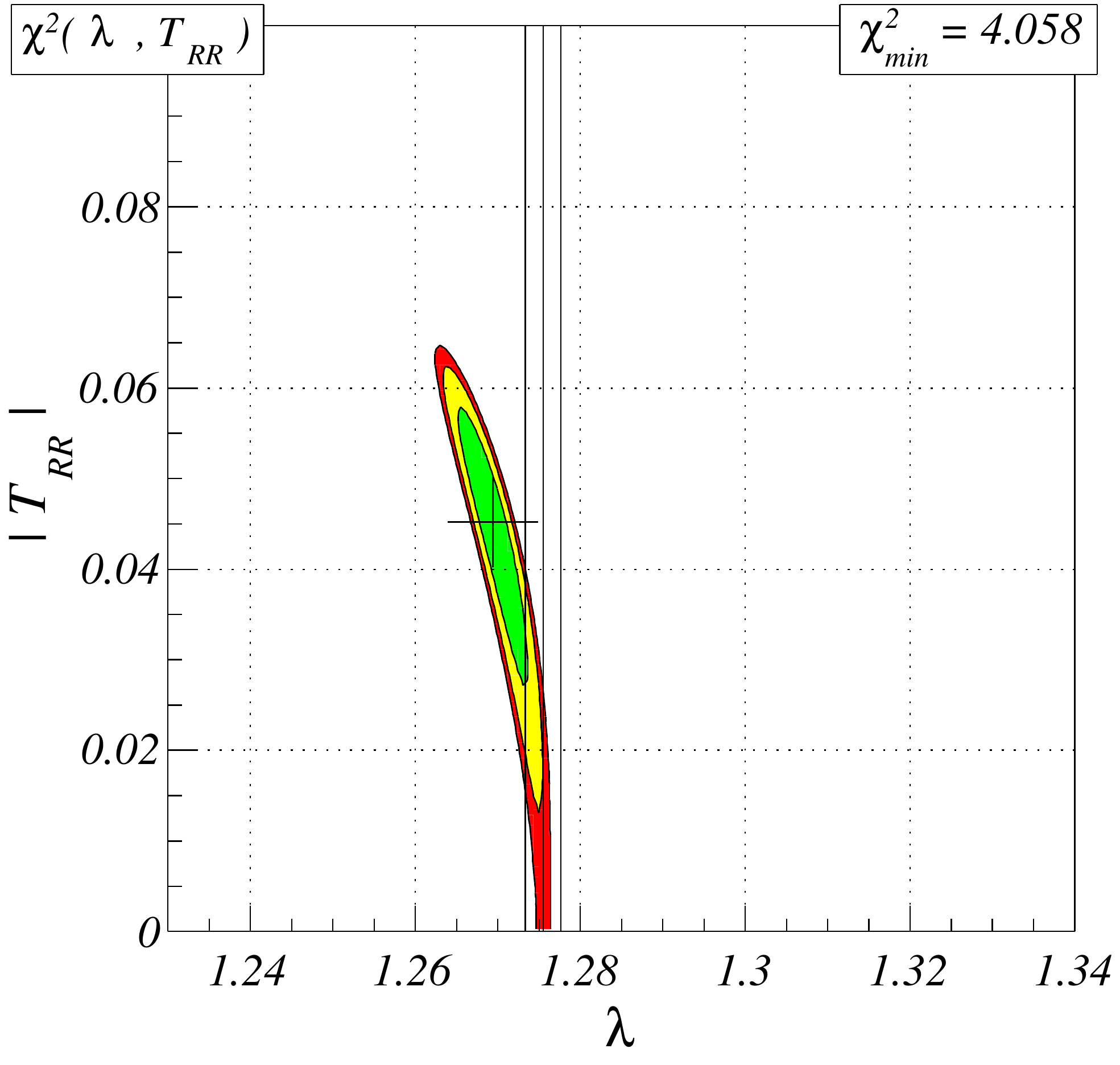}
\includegraphics[width=0.32\textwidth]{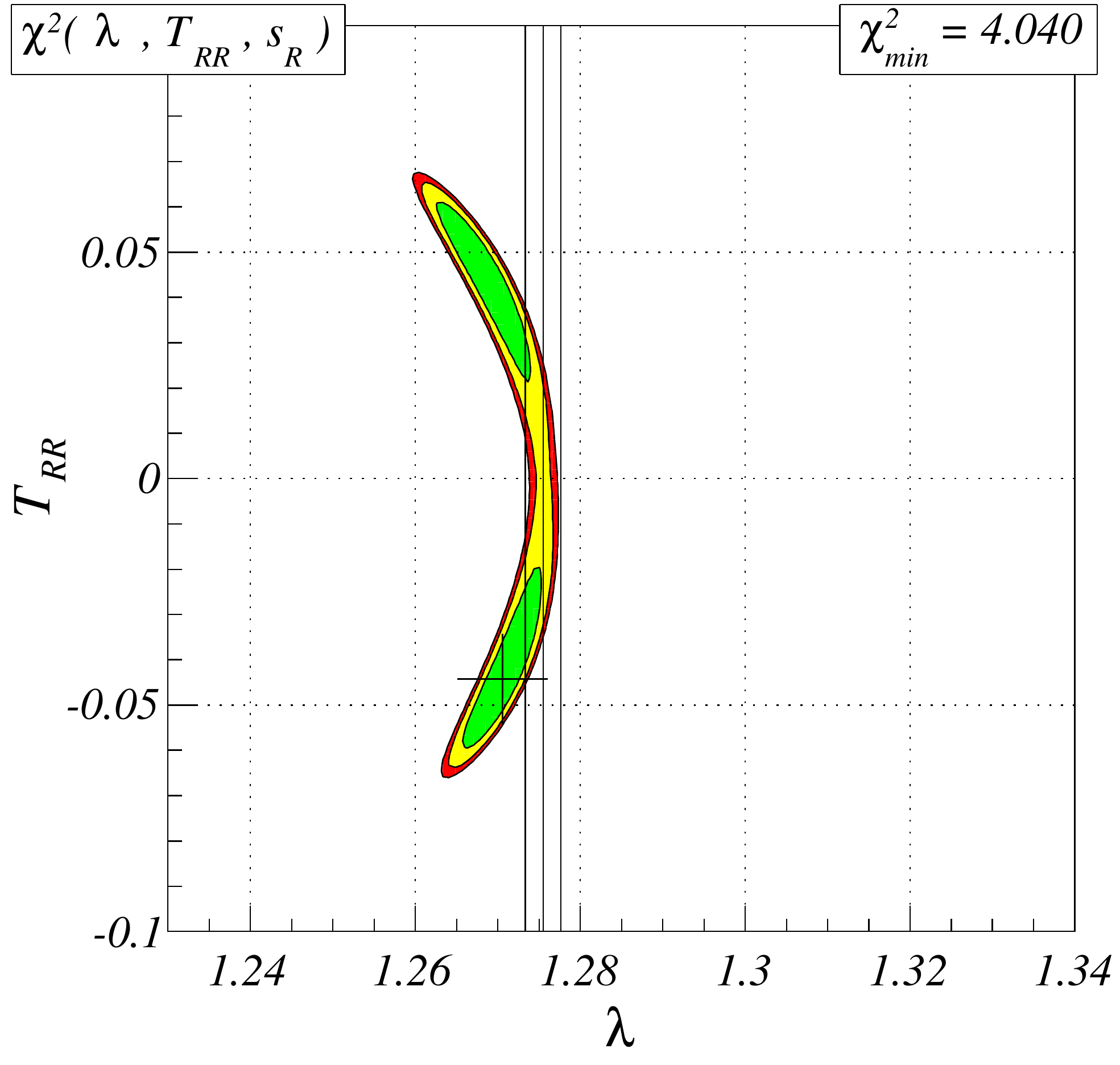}
\includegraphics[width=0.32\textwidth]{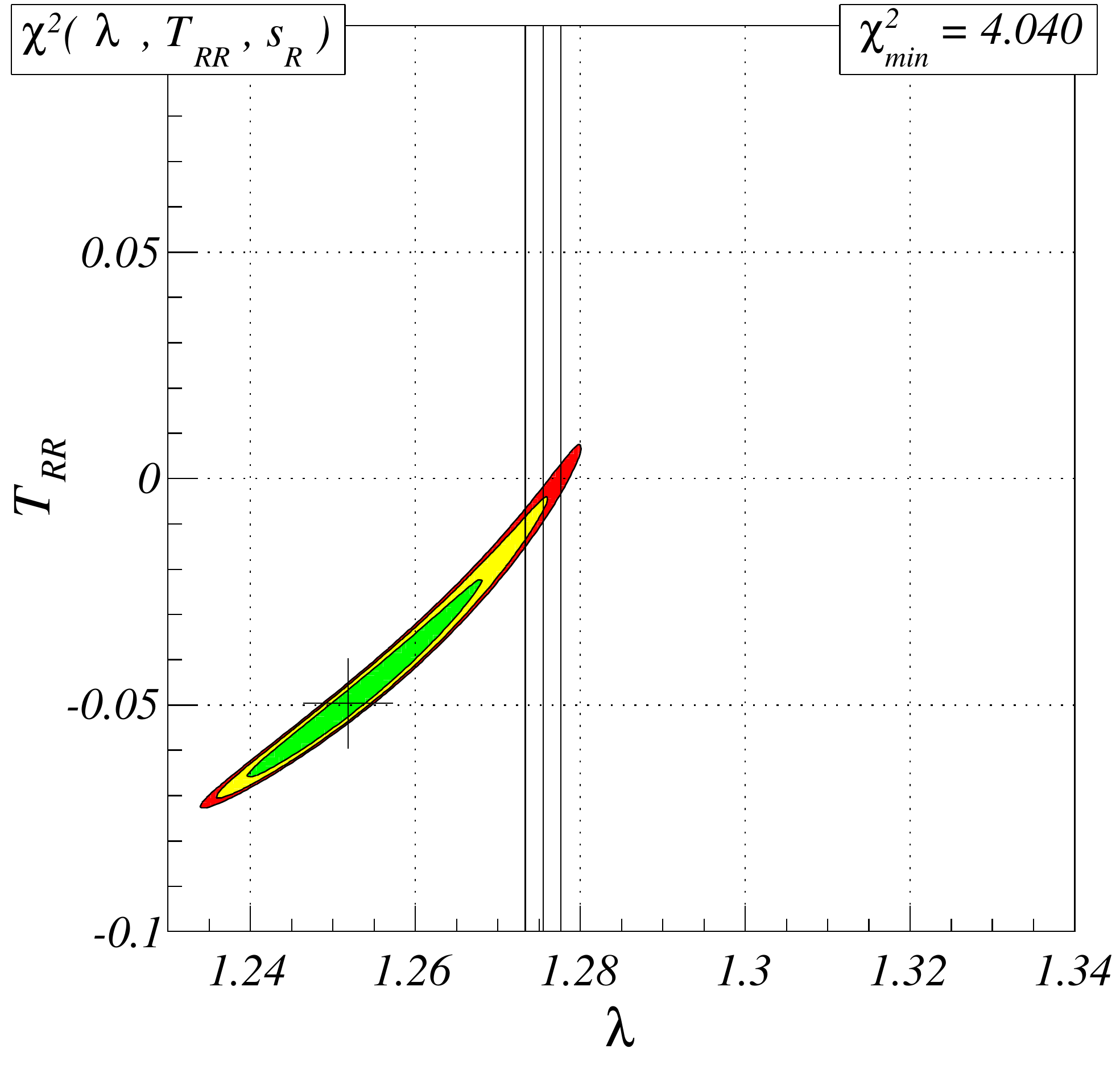}

\includegraphics[width=0.32\textwidth]{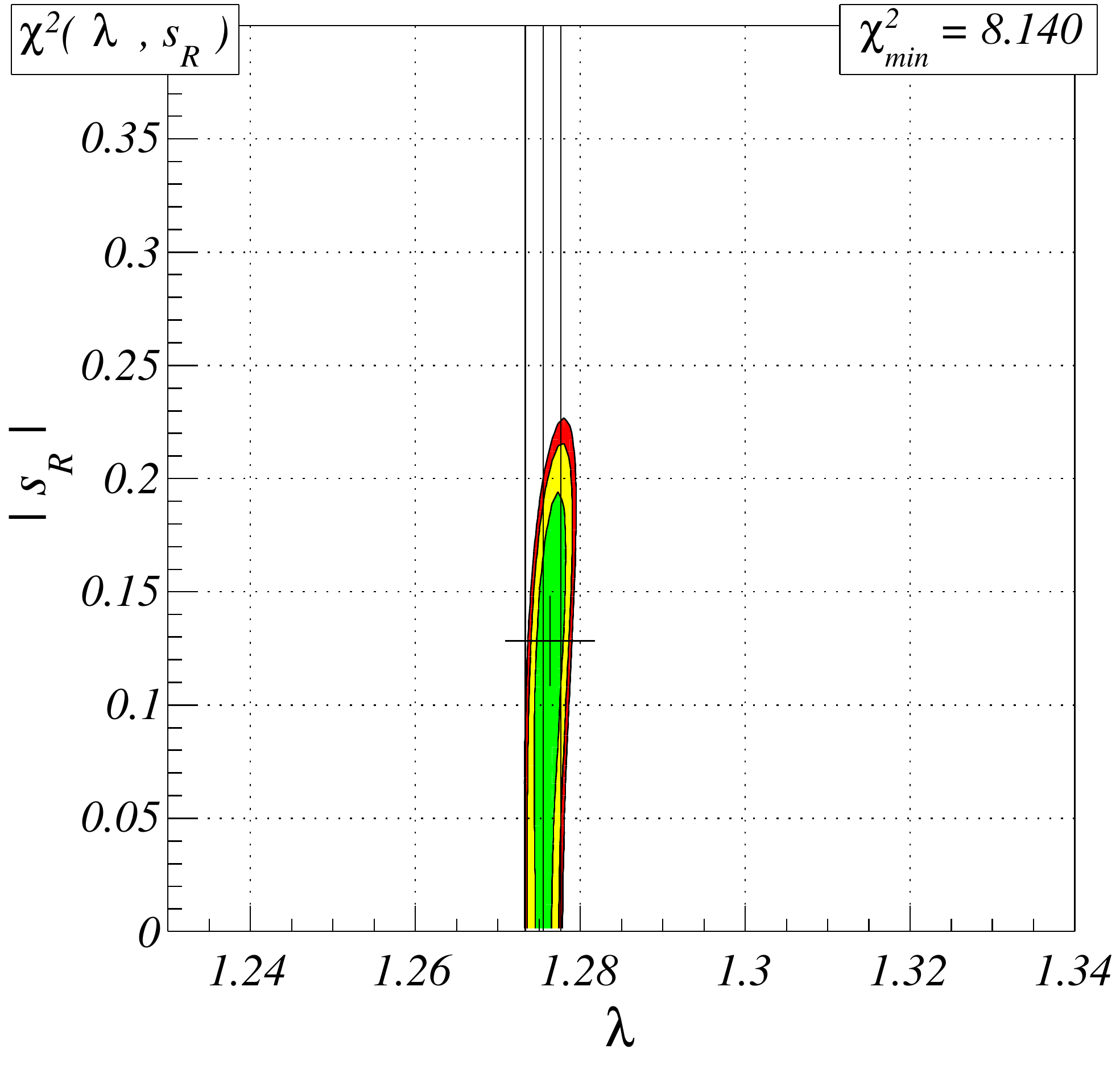}
\includegraphics[width=0.32\textwidth]{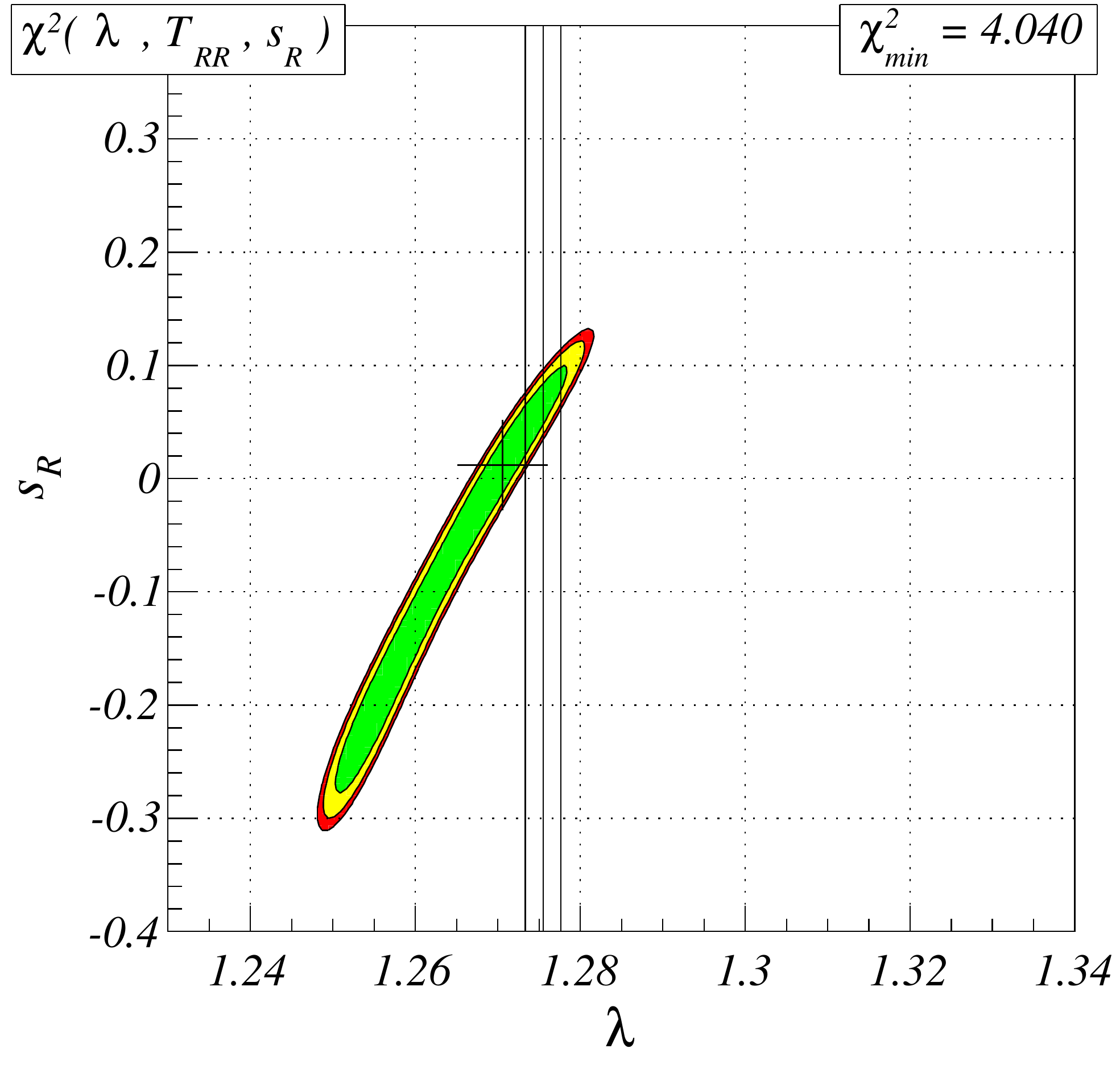}
\includegraphics[width=0.32\textwidth]{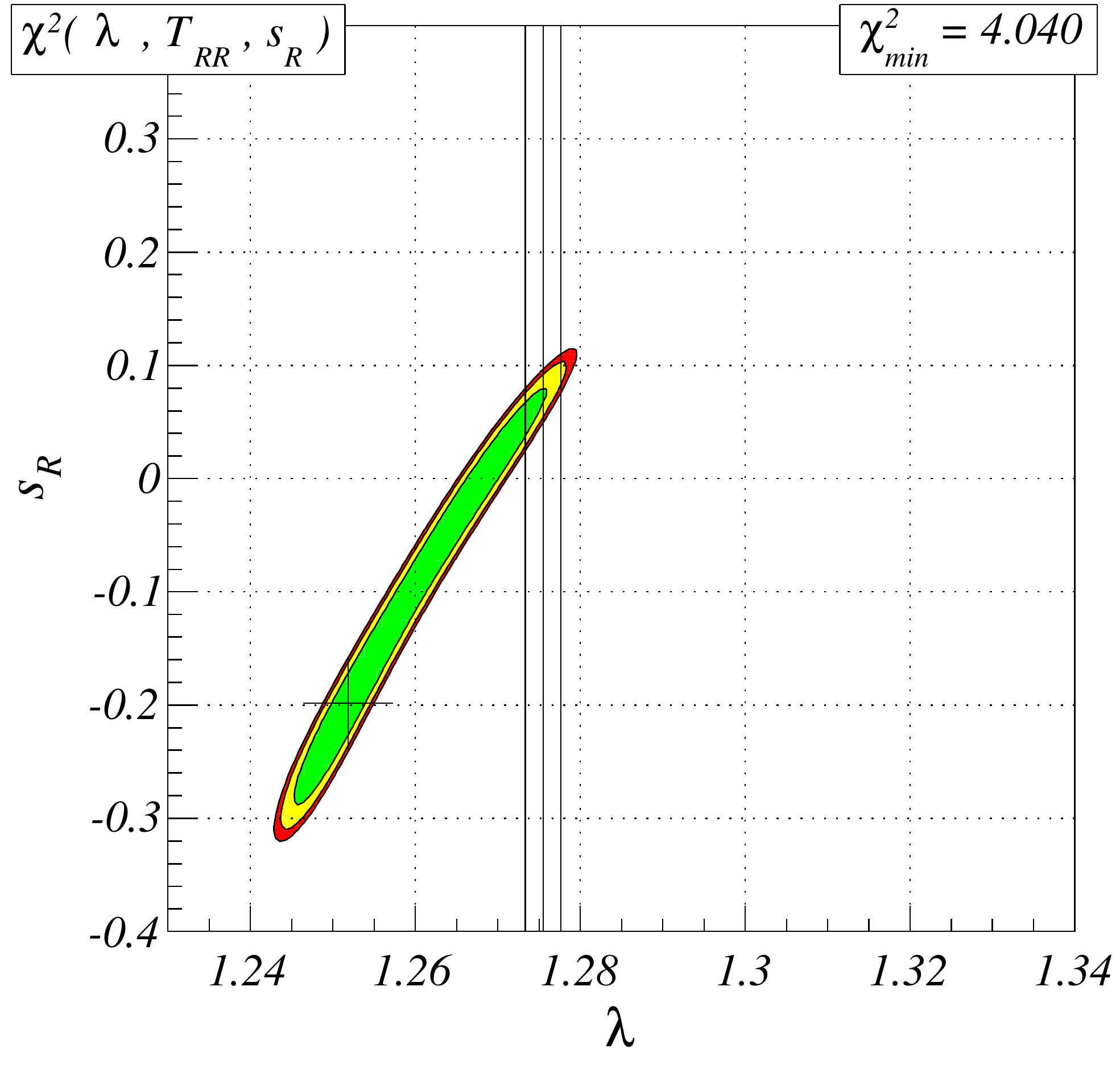}

\includegraphics[width=0.32\textwidth]{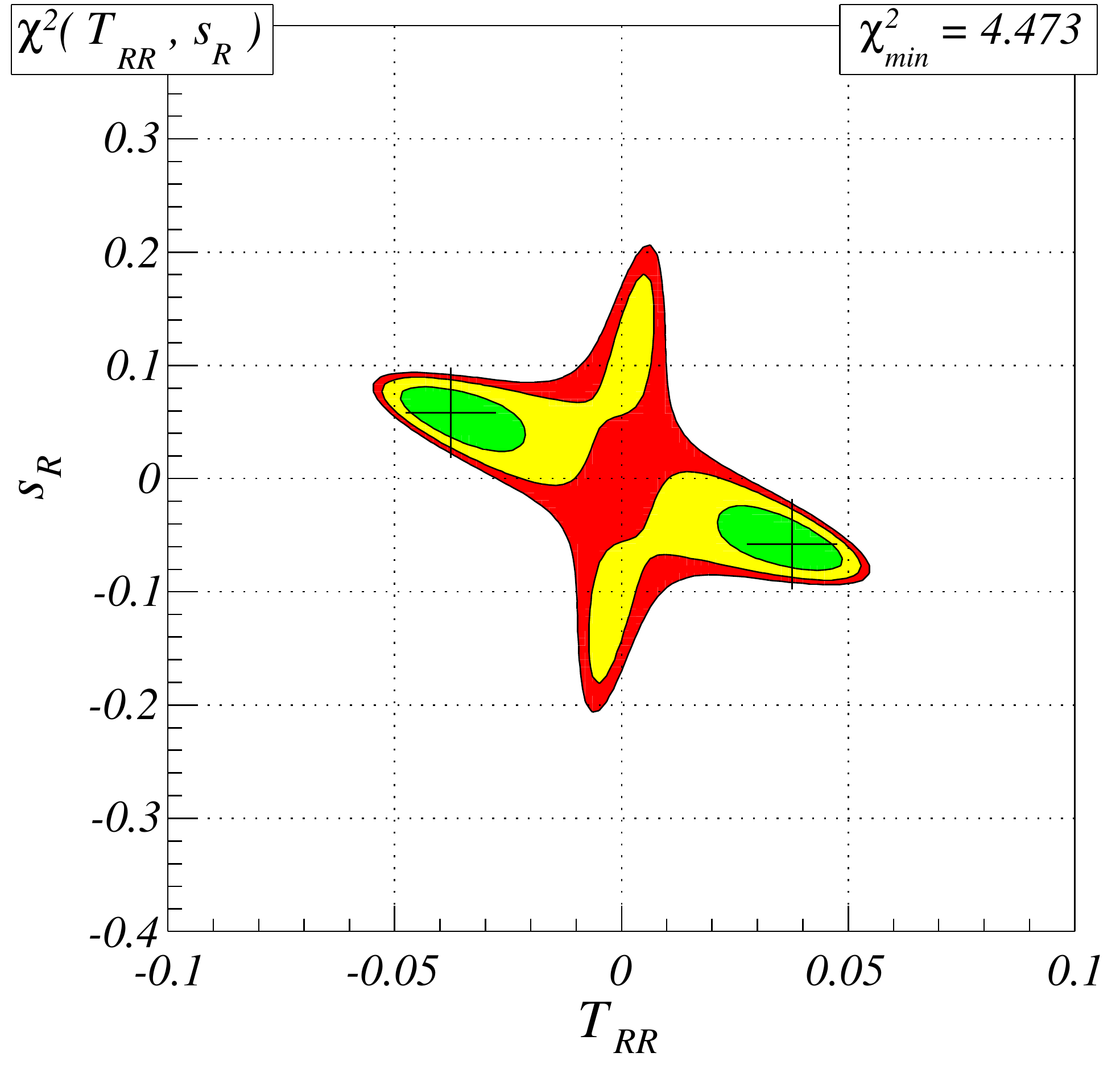}
\includegraphics[width=0.32\textwidth]{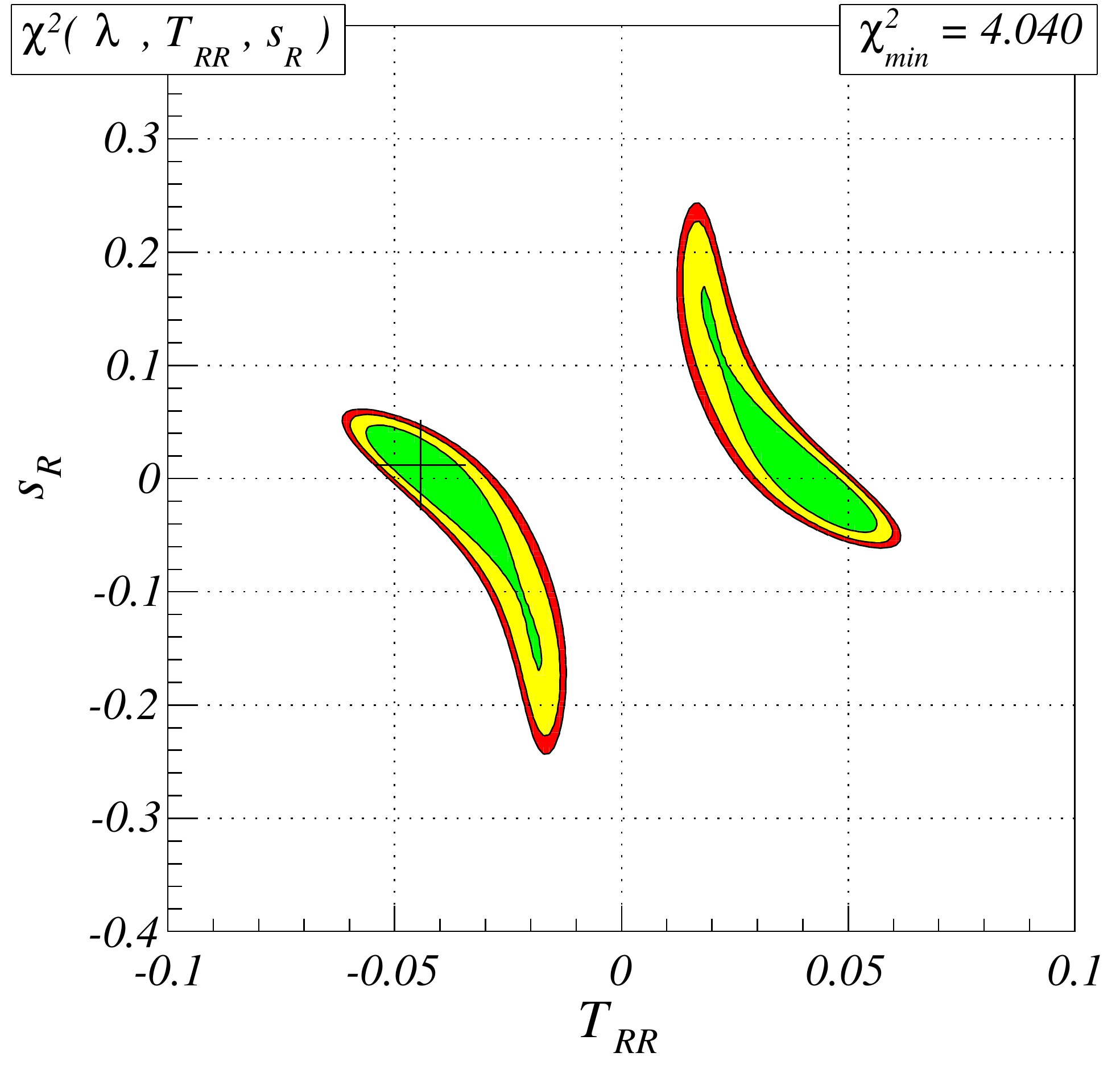}
\includegraphics[width=0.32\textwidth]{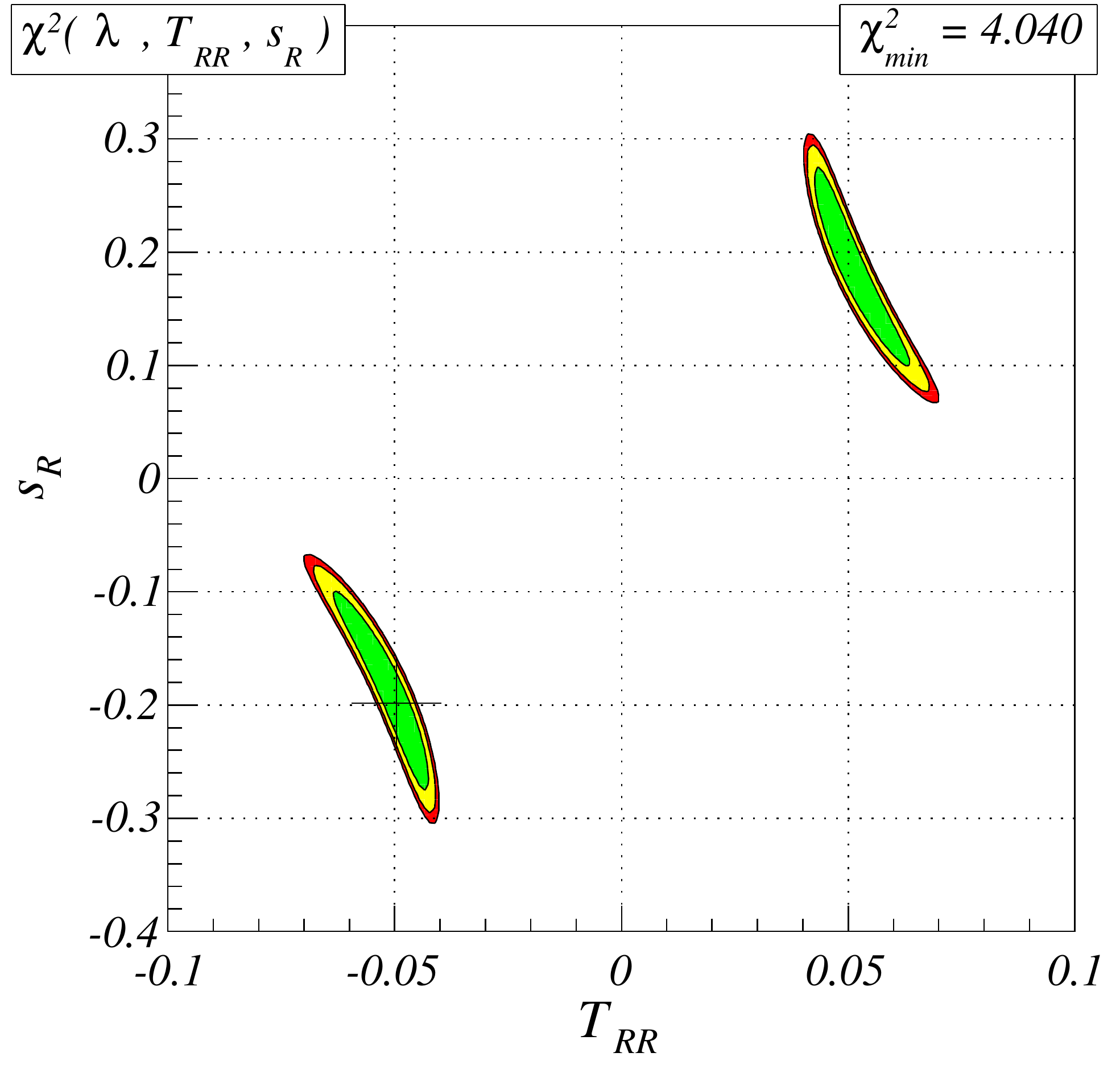}

\includegraphics[width=0.5\textwidth]{ContourLegend-eps-converted-to.pdf}

\caption{\label{T_RR_S_RR}The results of fits of right tensor and right scalar couplings to the data presented in the TABLE~\ref{data_table}. The arrangement of plots and description is analogical to that in FIG.~\ref{V_RL_V_RR}.}
\end{figure*}

\begin{figure*}[p!]
\centering

\includegraphics[width=0.32\textwidth]{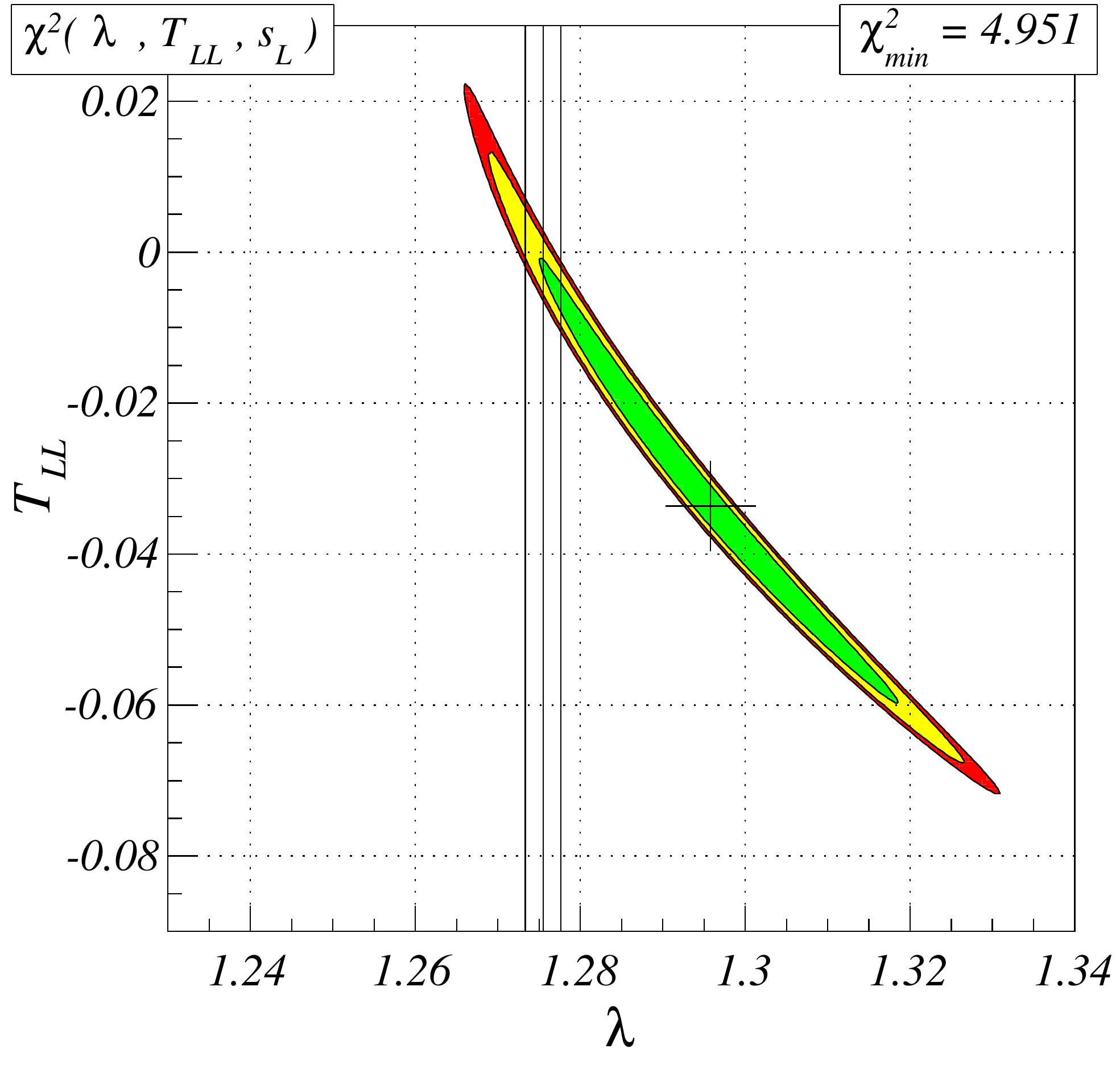}
\includegraphics[width=0.32\textwidth]{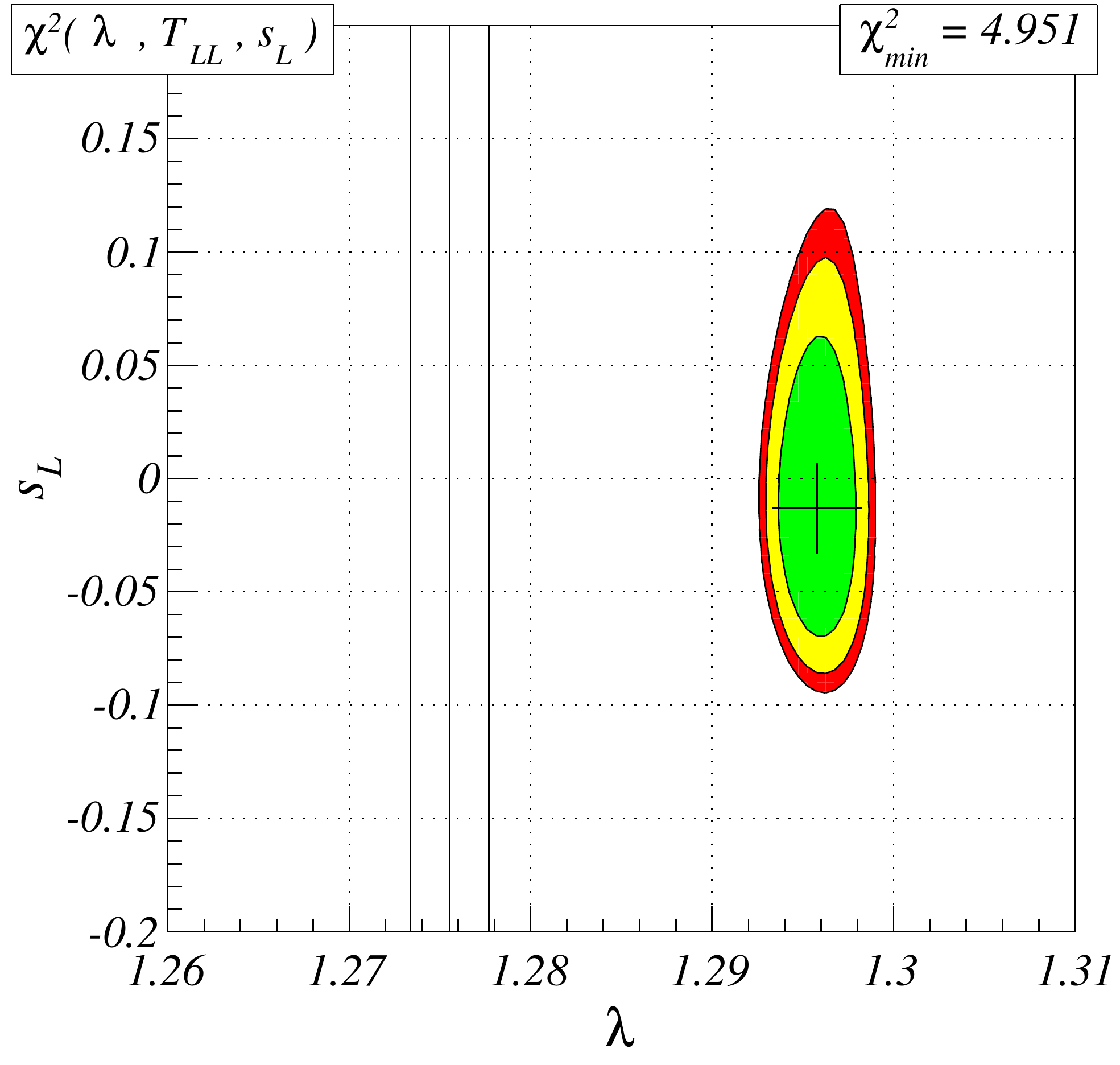}
\includegraphics[width=0.32\textwidth]{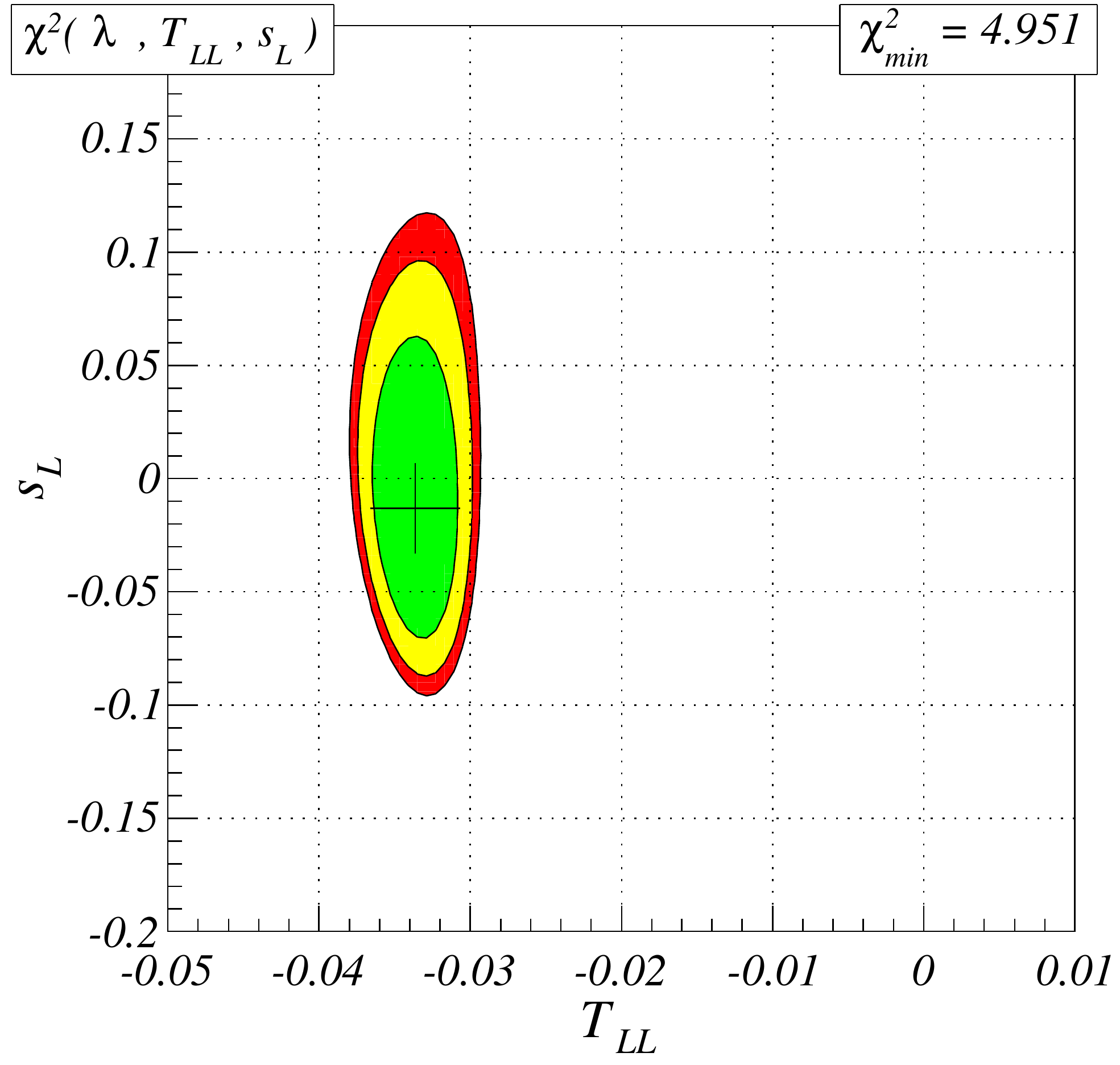}

\includegraphics[width=0.32\textwidth]{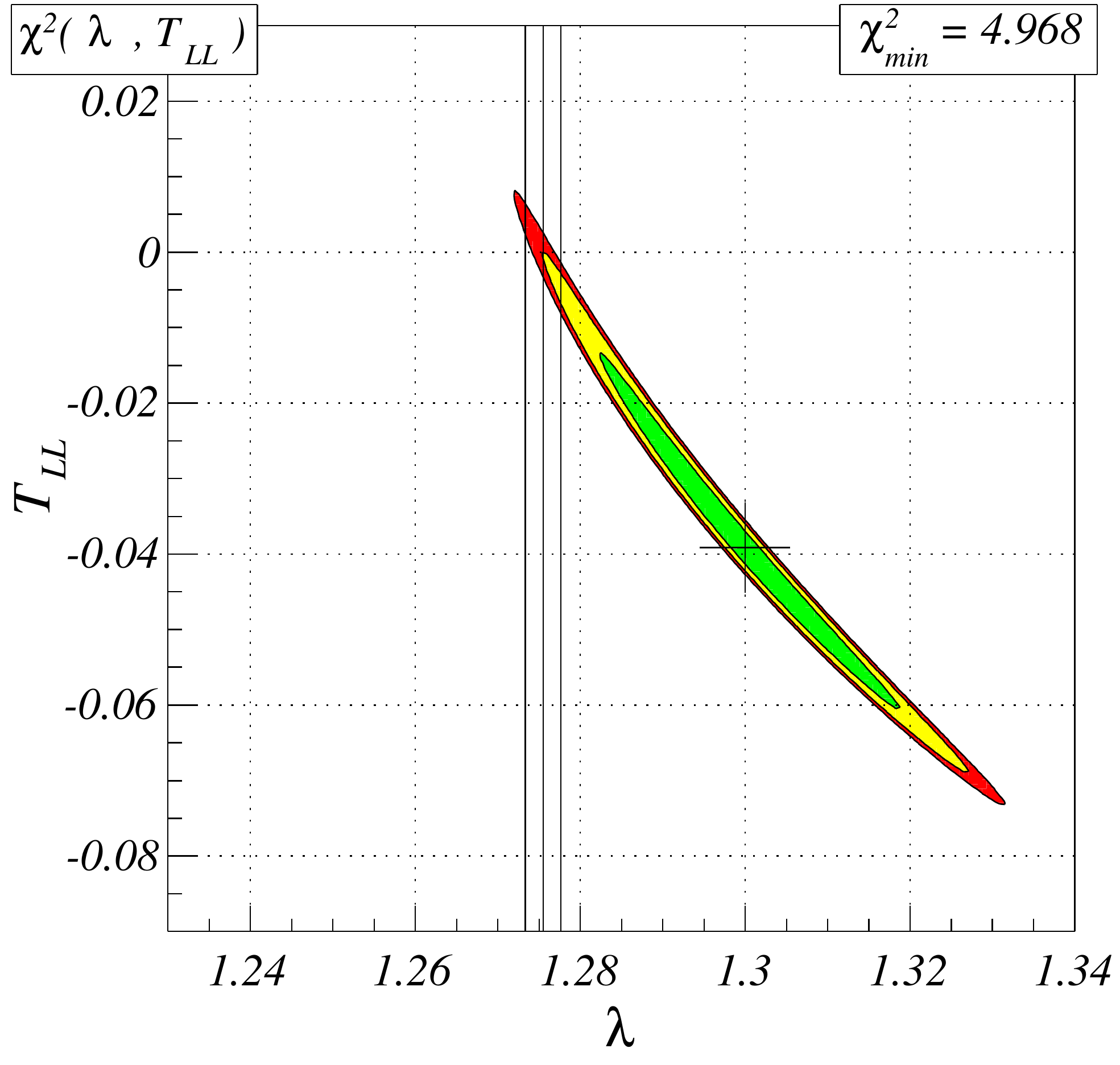}
\includegraphics[width=0.32\textwidth]{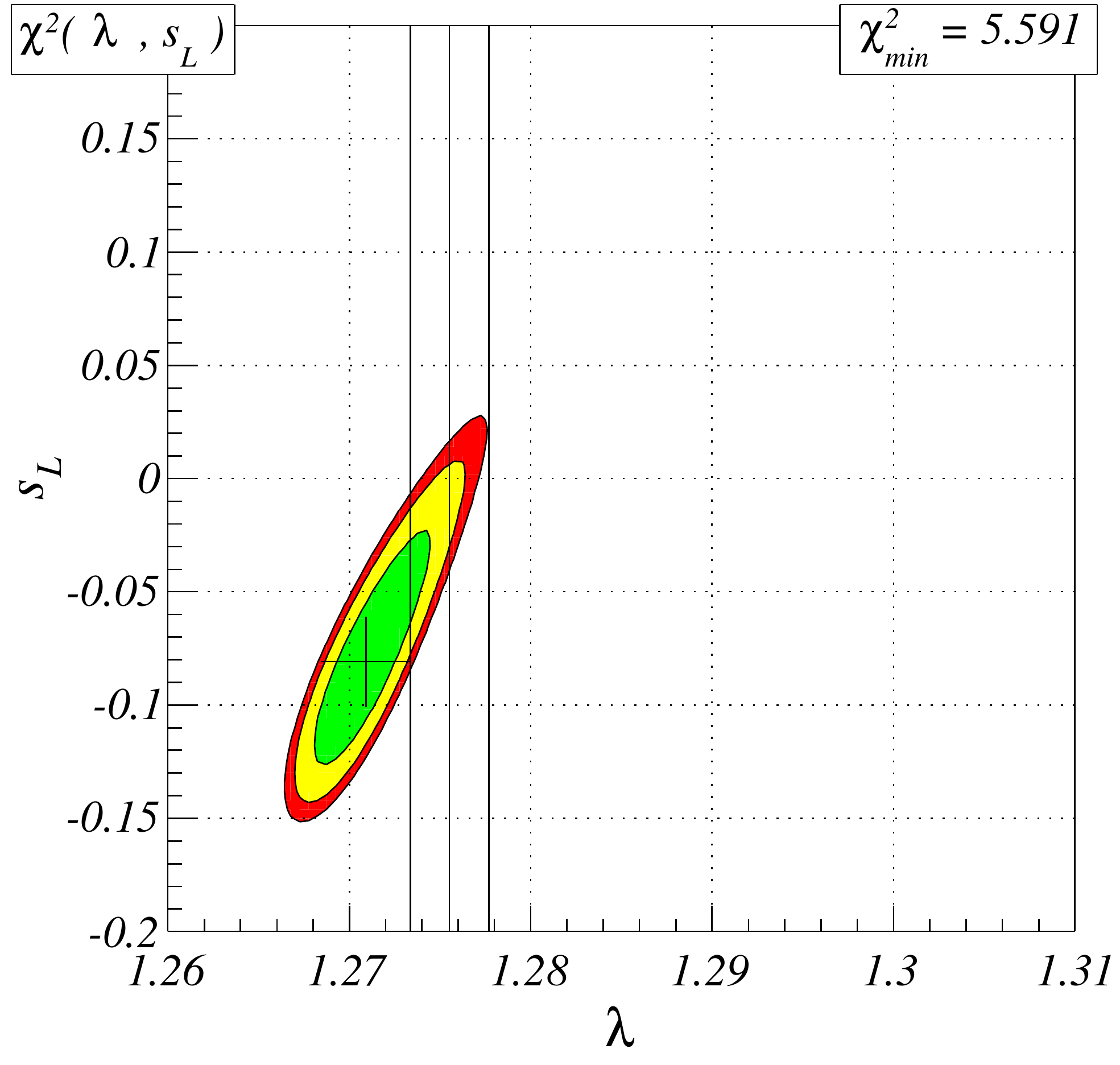}
\includegraphics[width=0.32\textwidth]{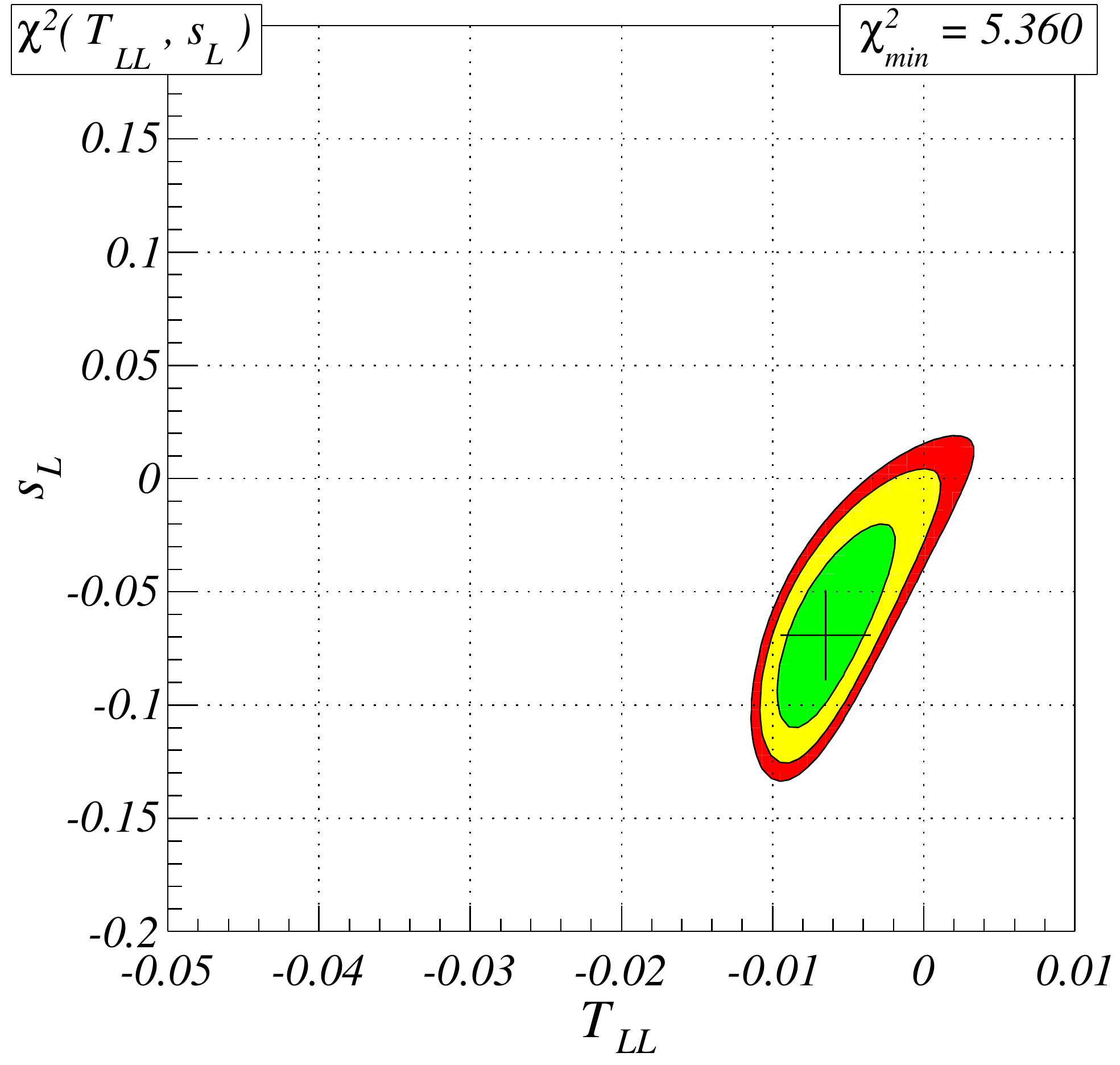}

\includegraphics[width=0.5\textwidth]{ContourLegend-eps-converted-to.pdf}

\caption{\label{T_LL_S_LL}The results of fits of left tensor and left scalar couplings solely to the data presented in the TABLE~\ref{data_table}. The first row groups slices of the three-dimensional $\chi^2$ volume of the respective three-parameter fit, while in the second we present results of two-parameter fits. As in FIG.~\ref{V_RL_V_RR} we list fitted parameters as the arguments of the $\chi^2$ function, we quote the value of the $\chi^2$ at the minimum and mark by the cross the position of the $\chi^2$ minimum. As before, the solid vertical lines mark the $95.45\%\ \textnormal{C.L.}$ interval on $\lambda$ (Eq.~(\ref{g_A_fit})).}
\end{figure*}

\begin{figure*}[p!]
\centering

\includegraphics[width=0.32\textwidth]{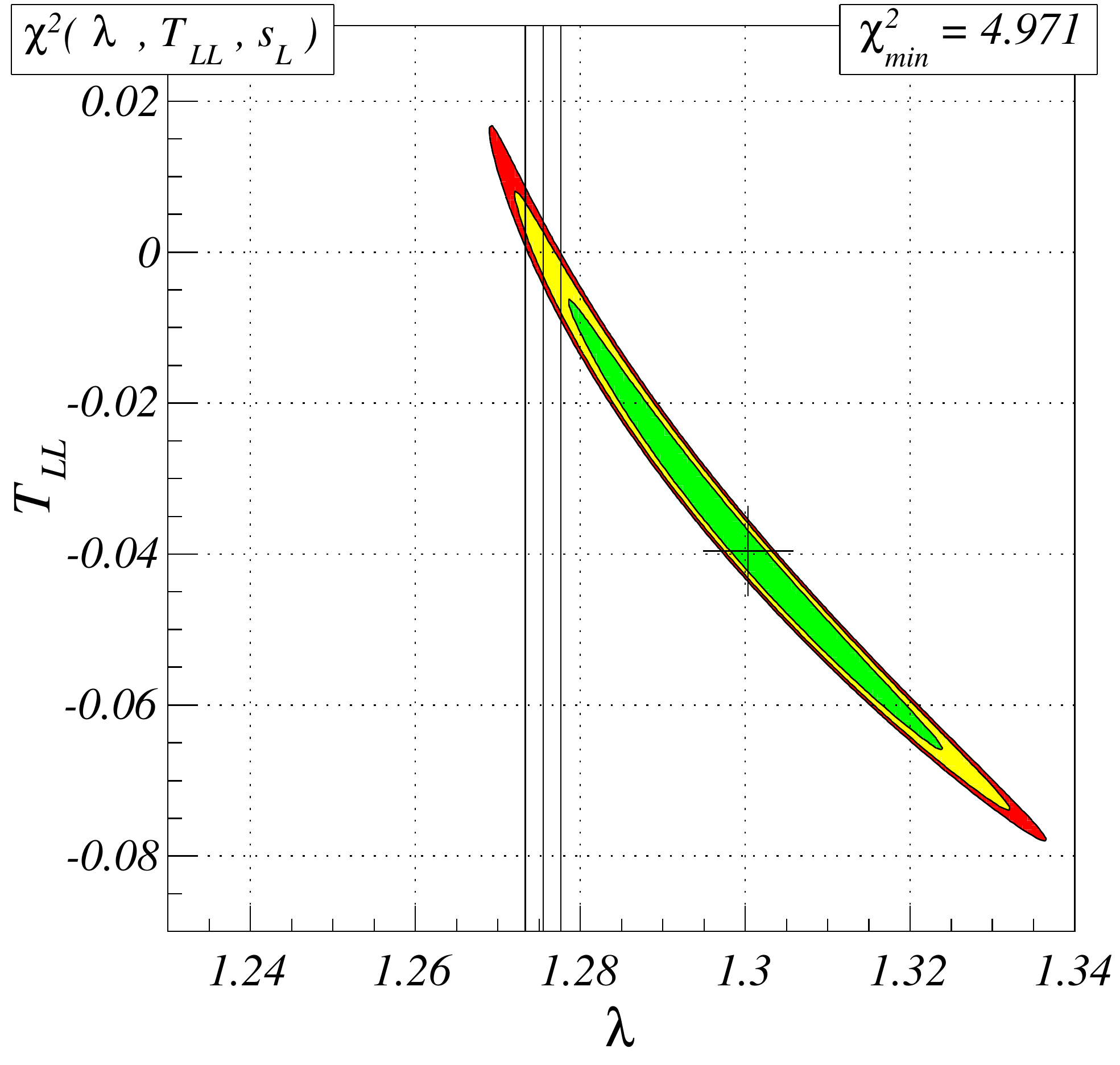}
\includegraphics[width=0.32\textwidth]{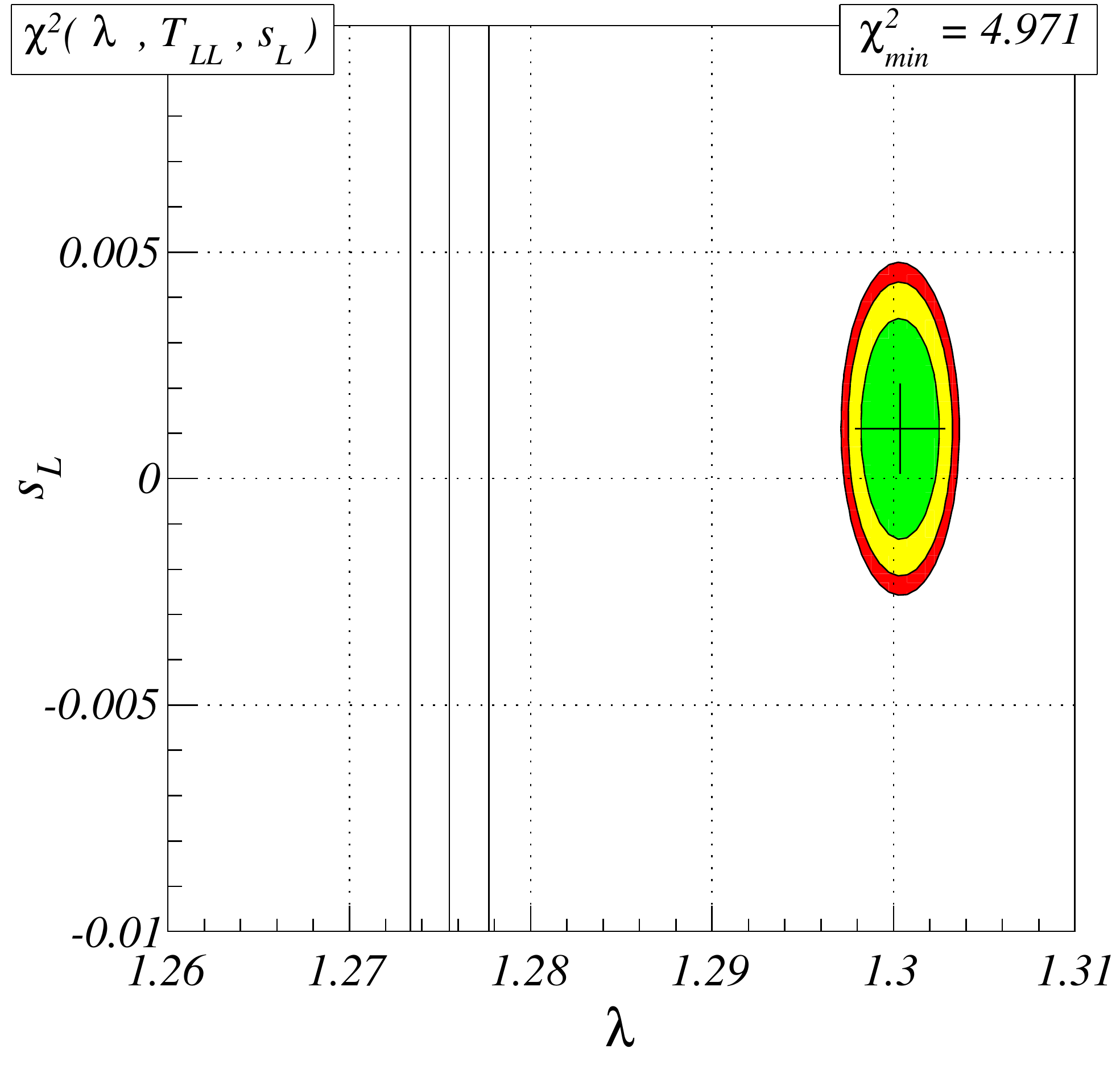}
\includegraphics[width=0.32\textwidth]{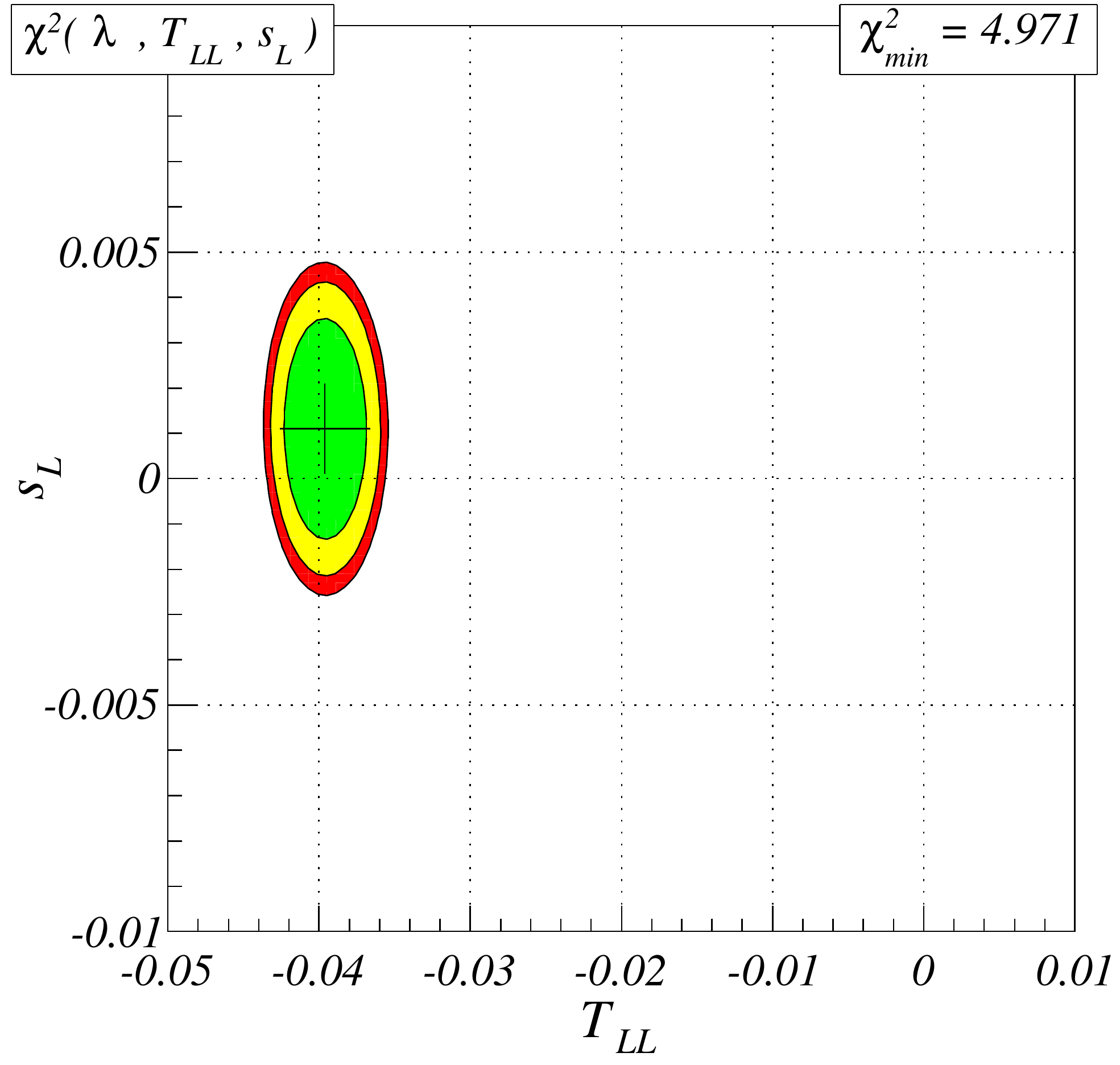}

\includegraphics[width=0.32\textwidth]{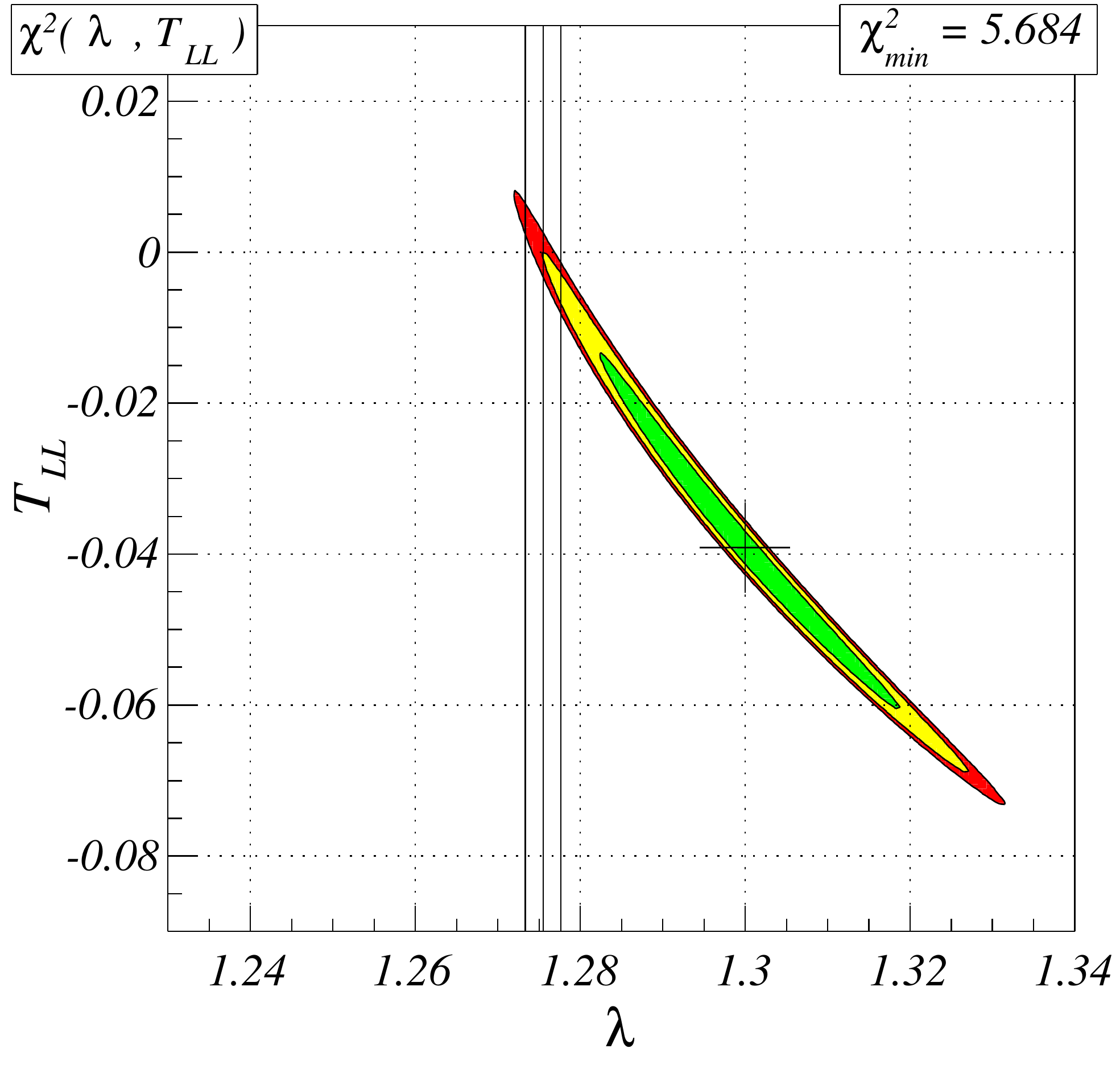}
\includegraphics[width=0.32\textwidth]{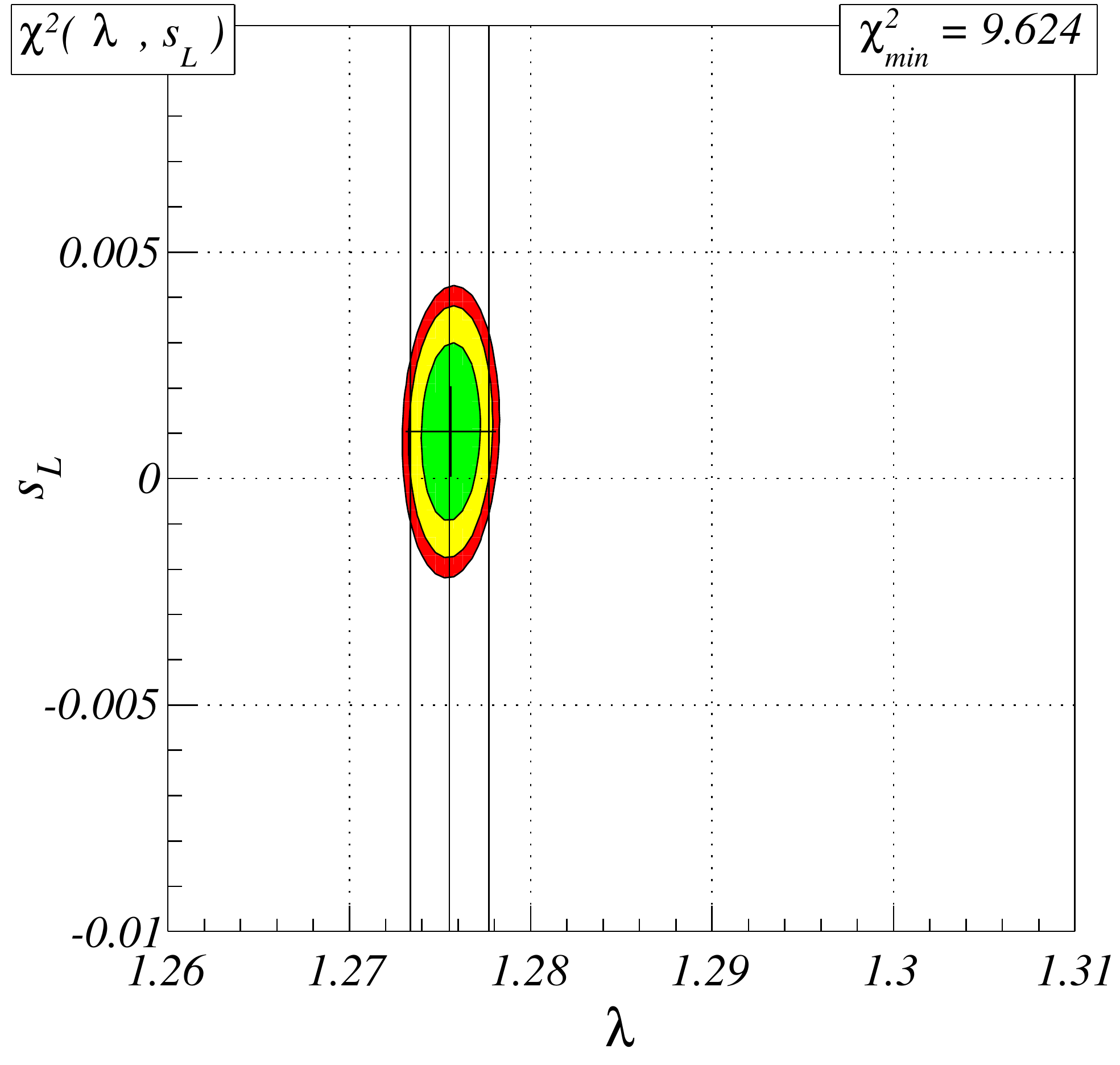}
\includegraphics[width=0.32\textwidth]{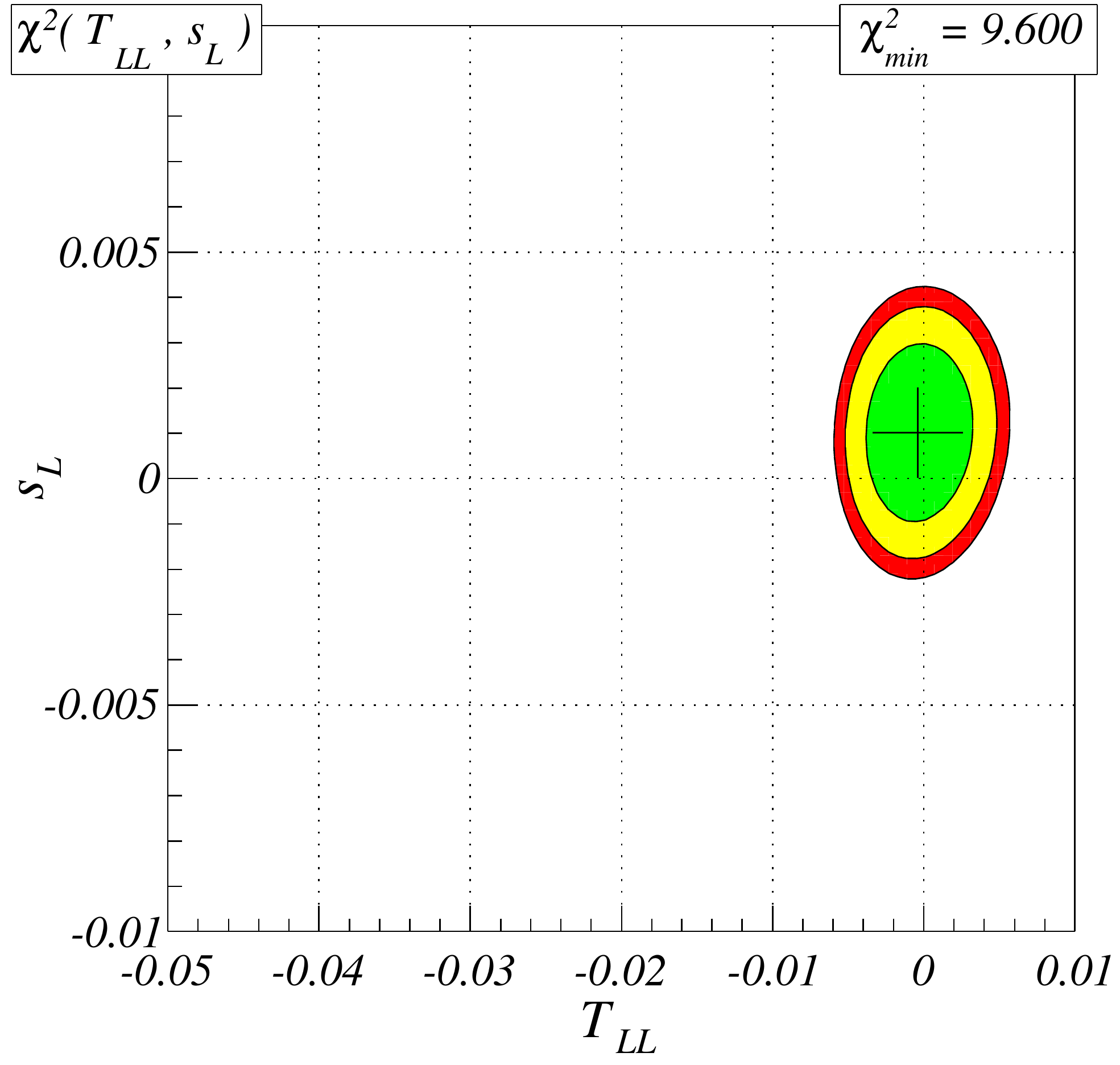}

\includegraphics[width=0.5\textwidth]{ContourLegend-eps-converted-to.pdf}

\caption{\label{T_LL_S_LL_bF}The results of fits of left tensor and left scalar couplings to the data presented in the TABLE~\ref{data_table} supplemented with the measurement of $b_F$. The arrangement of plots and description is analogical to that in FIG.~\ref{T_LL_S_LL}.}
\end{figure*}

Next we consider cases when one or two of $V_{RL}$, $V_{RR}$, $T_{kk}$, $s_k$ parameters are nonzero together with $\lambda$, that goes beyond its SM form as before. Results of such two and three-parameter fits are presented in FIGs.~\ref{V_RL_V_RR} -- \ref{T_LL_S_LL_bF}. Let us start with a few general remarks: (i)~we list only those arguments of $\chi^2$ that are actually fitted --- the rest are set to $0$ while $\lambda$, if not fitted, is set to its central value given in Eq.~(\ref{g_A_fit}), (ii)~the two-dimensional plots in the case of three-parameter fits are obtained by intersecting the corresponding three-dimensional $\chi^2$ volume with a plane that includes the three-parameter $\chi^2$ minimum point and is parallel to the respective planes spanned on the main axes in the parameter space. Enclosing general remarks we describe particular fits in more details below.

\begin{table*}
\centering
\begin{tabular}{l|rcl|rcl}
 \multicolumn{1}{c}{$\chi^2$} \vline &
 \multicolumn{3}{c}{without $b_F$} \vline &
 \multicolumn{3}{c}{with $b_F$} \\
\hline
 $ \chi^2 ( \lambda, T_{LL} ) $ &
 $ -0.022 < $ & $ b ( 0.1003 ) $ & $ < 0.19 $ &
 \multicolumn{3}{c}{the same as} \\
 &
 $ -0.020 < $ & $ b_\nu ( 0.0926 ) $ & $ < 0.17 $ &
 \multicolumn{3}{c}{without $b_F$} \\
\hline
 $ \chi^2 ( \lambda, s_L ) $ &
 $ -0.052 < $ & $ b ( -0.0276 ) $ & $ < 0.0096 $ &
 $ -0.00075 < $ & $ b ( 0.000352 ) $ & $ < 0.0015 $ \\
 &
 $ -0.066 < $ & $ b_\nu ( -0.0351 ) $ & $ < 0.013 $ &
 $ -0.00096 < $ & $ b_\nu ( 0.000449 ) $ & $ < 0.0019 $ \\
\hline
 $ \chi^2 ( s_L, T_{LL} ) $ &
 $ -0.024 < $ & $ b ( -0.00656 ) $ & $ < 0.011 $ &
 $ -0.015 < $ & $ b ( 0.00135 ) $ & $ < 0.016 $ \\
 &
 $ -0.037 < $ & $ b_\nu ( -0.0142 ) $ & $ < 0.0091 $ &
 $ -0.014 < $ & $ b_\nu ( 0.00137 ) $ & $ < 0.015 $ \\
\hline
 $ \chi^2 ( \lambda, s_L, T_{LL} ) $ &
 $ -0.28 < $ & $ b ( 0.0821 ) $ & $ < 0.58 $ &
 $ -0.045 < $ & $ b ( 0.1019 ) $ & $ < 0.20 $ \\
 &
 $ -0.28 < $ & $ b_\nu ( 0.0742 ) $ & $ < 0.58 $ &
 $ -0.041 < $ & $ b_\nu ( 0.0942 ) $ & $ < 0.18 $ \\
\end{tabular}
\caption{\label{b_table}The maximal and minimal values of $b$ and $b_\nu$ calculated from parameter combinations at $95.45\%\ \textnormal{C.L.}$. The values in parentheses correspond to $\chi^2$ minima.}
\end{table*}

First, we made fits to the neutron data presented in the TABLE~\ref{data_table} in all possible two and three parameter combinations in which $b \equiv 0$ and $b_\nu \equiv 0$. These results can be divided into two main groups: (i)~the fits when the only nonzero parameters are $V_{RL}$ or $V_{RR}$ or both of them simultaneously (vector couplings fits) --- see FIG.~\ref{V_RL_V_RR}, (ii)~the fits when the only nonzero parameters are $s_R$ or $T_{RR}$ or both of them simultaneously (right tensor and right scalar couplings fits) --- see FIG.~\ref{T_RR_S_RR}. Combining vector couplings (first group) with right tensor or right scalar couplings (second group) results in $b$ and $b_\nu$ being not identically zero. In the case of two-parameter fits there are $2$ equivalent minima because of the $\chi^2$ symmetry given in Eq.~(\ref{chi2_symmetry}). In the case of three-parameter fits there are $2 \times 2 = 4$ equivalent minima: the minimization procedure found two equivalent minima corresponding to the different values of $\lambda$ and, for each of these values, we have two sets of $V_{RL}$, $V_{RR}$ or $s_R$, $T_{RR}$ parameters from $\chi^2$ symmetry in Eq.~(\ref{chi2_symmetry}).

Finally, we made fits when the only nonzero parameter is $s_L$ or $T_{LL}$ as well as both of them simultaneously (left tensor and left scalar couplings) using only neutron data as before --- these results are presented in FIG.~\ref{T_LL_S_LL}. There is only one minimum in each fit and in all cases $b$ as well as $b_\nu$ are not identically zero. Moreover, we report that the calculated values of $b$ and $b_\nu$ are rather big (see \eg \cite{Dubbers_Schmidt}) especially when all parameters are fitted together as it can be seen in the TABLE~\ref{b_table} --- this causes strong dependency of the results on the $\langle W^{-1}\rangle_i$ values. To reduce this effect we added to the data set a measurement of $b$ in superallowed Fermi decays \cite{Hardy_Towner} 
\begin{equation}
b_F = -0.0022 \pm 0.0026.
\end{equation}
We therefore extended our previous $\chi^2$ in Eq.~(\ref{chi2}) with an extra term
\begin{equation}
\chi^2 \longrightarrow \chi^2 + \frac{(b_F + 0.0022)^2}{(0.0026)^2},
\end{equation}
where the general formula for $b_F$ in given by\footnote{The formula for $b_F$ can easily be derived from formulas obtained in Ref.~\cite{Jackson}, after appropriate change of the parametrization in the form given in Ref.~\cite{Herczeg}. We assume that in the case of nuclear decays, that were studied in Ref.~\cite{Hardy_Towner}, to obtain the quoted $b_F$ value the same anti-neutrino states are kinematically allowed as in the case of the free neutron beta decay.} 
\begin{equation} \label{b_F}
b_F = \frac{-2 \left[s_R (V_{RR}+ V_{RL})+s_L \right]}{(V_{RR} +V_{RL})^2+s_R^2+s_L^2+1}.
\end{equation}
These results are presented in FIG.~\ref{T_LL_S_LL_bF} and the calculated values of $b$ and $b_\nu$ are listed in the TABLE~\ref{b_table}. From Eq.~(\ref{b_F}) we see that in the case under consideration (\ie when $V_{RL} = 0$, $V_{RR} = 0$, $s_R = 0$ and $T_{RR} = 0$) $b_F = -2s_L/(1 + s_L^2)$. Thus, including the experimental value for $b_F$ really affects only fits in which $s_L$ is one of the nonzero parameters, while $\chi^2(\lambda, T_{LL})$ is affected just by an overall shift of $(0.0022)^2/(0.0026)^2 \approx 0.716$ with respect to the same $\chi^2$ calculated without $b_F$. 
When comparing FIG.~\ref{T_LL_S_LL_bF} with plots in FIG.~\ref{T_LL_S_LL}, the limits on $s_L$ are stronger and the ones on $T_{LL}$ are practically unchanged.

It is difficult to compare our results with the recent ones in  Refs.~\cite{Severijns,Konrad} because of different parametrizations. If we set $a_{LR} \equiv 0$ we can easier compare our results with the analysis in Ref.~\cite{Herczeg}. When only one beyond SM parameter is fitted together with $\lambda = g_A/g_V$ to free neutron data we conclude that our constrains are not as strong as those coming from joint analysis where also pion and nuclear beta decays are included. 

We would like to point out that, in cases when $\lambda$ is fitted together with only one parameter beyond the Standard Model, there is no significant difference with respect to our previous results \cite{our_acta_2011} except for a shift in $\lambda$, caused by the different selection of $A$ decay parameter measurements. As expected, the change of the data selection influences the result of the one--parameter fit on $\lambda$.

In the case of three parameter fits with $\chi^2(\lambda,V_{LR},V_{RR})$ and $\chi^2(\lambda,s_{R},T_{RR})$
we obtained four equivalent minima. Two of them give the $\lambda$ which differs from the SM value at the level of $3 \sigma$, and the next two where the difference is much larger. This discrepancy is also large in three parameter fits with $\chi^2(\lambda,s_L,T_{LL})$.

\section{\label{sec:conclusions}Conclusions}

We found limits on parameters describing physics beyond the Standard Model in the neutron beta decay including the last experimental data. The fits were done using the parametrization in which the influence of New Physics at the quark--lepton level can be easily separated from the nucleon structure in the vector part of the interaction. Our result for one parameter SM fit of $\lambda$ slightly differs from that of PDG because of different data selection. In the presence of New Physics one can find values of $\lambda$ at the $\chi^2$ minimum that are relatively close to that in the SM and other which deviate significantly form the SM expectations. We confirm what was already mentioned in the literature that free neutron decay alone cannot give us the strongest limits for exotic couplings. Therefore we are looking forward for new, even more precise experimental data, especially of the still unknown values of $b$ and $b_\nu$ coefficients.

\begin{acknowledgments}

Micha\l{}~Ochman acknowledges a scholarship from the \'S{}WIDER project co-financed by the European Social Fund.
We would also like to acknowledge the usage of the Maxima computer algebra system \cite{MAXIMA} for symbolic calculations and the ROOT framework \cite{ROOT} for the numerical part (and the graphical presentation of the results).

\end{acknowledgments}


\appendix*

\section{Formulas for correlation coefficients}

Below we present the formulas for the correlation coefficients $a$, $b$, $A$, $B$ ($B = B_{0} + b_\nu m_e / E_e$) from Eqs.~(\ref{d_Gamma},\ref{B_coeff}) as functions of the parameters defined in Eqs.~(\ref{VST_parameters},\ref{akl_def},\ref{s_L_R}).

\begin{eqnarray}
\xi 
& = & 3 \lambda^2 \left[(V_{RR}-V_{RL})^2 +1\right] \nonumber \\ 
&   & +   (V_{RR}+V_{RL})^2 +1 \nonumber \\
&   & +   12(T_{RR}^2+T_{LL}^2)+s_R^2+s_L^2 \\   
&   & \nonumber \\ 
a \xi 
& = & -  \lambda^2  \left[(V_{RR}-V_{RL})^2 +1\right] \nonumber \\   
&   & +   (V_{RR}+V_{RL})^2 +1 \nonumber \\
&   & +   4 (T_{RR}^2+T_{LL}^2)-s_R^2-s_L^2 \\   
&   & \nonumber \\ 
b \xi 
& = & - 12 \lambda \left[T_{RR} (V_{RR}-V_{RL})+T_{LL} \right] \nonumber \\   
&   & +   2 \left[s_R (V_{RR}+V_{RL})+s_L \right] \\   
&   & \nonumber \\ 
A \xi
& = & -  2 \lambda^2 \left[1 - (V_{RR}-V_{RL})^2 \right] \nonumber \\   
&   & +  2 \lambda \left[1 - \left(V_{RR}+V_{RL} \right) \left(V_{RR}-V_{RL} \right) \right] \nonumber \\   
&   & -  4 \left(2 T_{RR}^2+s_R T_{RR}-2 T_{LL}^2-s_L T_{LL}\right) \\   
&   & \nonumber \\ 
B_0 \xi
& = & 2 \lambda^2 \left[1 - (V_{RR}-V_{RL})^2 \right] \nonumber \\   
&   & +  2 \lambda \left[1 - \left(V_{RR}+V_{RL} \right) \left(V_{RR}-V_{RL} \right) \right] \nonumber \\
&   & -  4 \left(2 T_{RR}^2-s_R T_{RR}-2 T_{LL}^2+s_L T_{LL}\right) \\   
&   & \nonumber \\ 
b_\nu \xi
& = & 2 \lambda \left[(4 T_{RR}-s_R)(V_{RR}-V_{RL})
                 -(4 T_{LL}-s_L) \right] \nonumber \\   
&   & +   4 \left[T_{RR} (V_{RR}+V_{RL})-T_{LL} \right]
\end{eqnarray}

\end{document}